\documentclass[twocolumn,amssymb,secnumarabic, amsmath,nofootinbib]{revtex4}

\usepackage{graphicx}
\usepackage{amssymb}
\usepackage{booktabs}
\usepackage{color}

\newcommand{\ra}[1]{\renewcommand{\arraystretch}{#1}}
\newcommand{\rb}[1]{\renewcommand{\tabcolsep}{#1}}

\newcommand{\ord}{\mathcal{O}}
\newcommand{\nn}{\noindent}
\newcommand{\nb}{\nonumber}
\newcommand{\arccot}{{\rm arccot}}

\newcommand{\eq}[1]{\begin{equation} #1 \end{equation}}
\newcommand{\eqa}[1]{\begin{eqnarray} #1 \end{eqnarray}}
\newcommand{\C}[1]{{\cal C}_{#1}}
\newcommand{\Cp}[1]{{\cal C}_{#1'}}
\newcommand{\delC}[1]{\delta {\cal C}_{#1}}
\newcommand{\dC}[1]{{\cal C}_{#1}^{\rm NP}}
\newcommand{\dCp}[1]{{\cal C}_{#1^\prime}^{\rm NP}}
\newcommand{\av}[1]{\langle #1 \rangle}
\newcommand{\afb}{A_{\rm FB}}
\newcommand{\mpg}[1]{\begin{minipage}{\linewidth} \begin{center} #1 \end{center} \end{minipage}}

\newcommand{\sect}[1]{\section{\hspace{-0.3cm} #1}}

\begin{document}

\title{Understanding the $B \to K^* \mu^+\mu^-$ Anomaly}

\author{S\'ebastien Descotes-Genon$^{a}$, Joaquim Matias$^{b}$ and Javier Virto$^{b}$
\vspace{0.3cm}}

\affiliation{
\mpg{\mbox{$^{a}$Laboratoire de Physique Th\'eorique, CNRS/Univ. Paris-Sud 11 (UMR 8627), 91405 Orsay Cedex, France}\\ \vspace{0.1cm}}
\mbox{$^{b}$Universitat Aut\`onoma de Barcelona, 08193 Bellaterra, Barcelona}\\}

\begin{abstract}

We present a global analysis of the $B \to K^* (\to K \pi) \mu^+\mu^-$ decay using the recent LHCb measurements of the primary observables $P_{1,2}$ and $P_{4,5,6,8}^\prime$. 
Some of them exhibit large deviations with respect to the SM predictions. 
We explain the observed pattern of deviations through  a large New Physics contribution to the Wilson coefficient of the semileptonic operator $\ord_9$.
 This contribution has an opposite sign to the SM one, i.e.,  reduces the size of this coefficient significantly. 
 A good description of data is achieved  by allowing for New Physics contributions to the Wilson coefficients  $\C7$ and $\C9$ only.  We find a 4.5$\,\sigma$ deviation with respect to the SM prediction, combining the large-recoil $B \to K^* (\to K \pi) \mu^+\mu^-$ observables with other radiative processes. Once low-recoil observables are included the significance gets reduced to 3.9$\,\sigma$. We have tested different sources of systematics, none of them modifying our conclusions significantly.
 Finally, we propose additional ways of measuring the primary observables through new foldings.

\end{abstract}


\maketitle


The four-body $B \to K^*(\to K \pi) \mu^+\mu^-$ decay and its plethora of different observables~\cite{kruger,matias1,Lunghi:2006hc,buras,matias2,1202.4266,becirevic,bobeth,tensors,Alok:2010zd,Alok:2011gv, Altmannshofer:2011gn,Altmannshofer:2012az,Beaujean:2012uj,1207.2753} is becoming one of the key players not only in our search for New Physics (NP) in the flavour sector but also to guide us in the construction of viable new models, which explains the remarkable experimental effort devoted to its precise measurement~\cite{1108.0695,Wei:2009zv,Lees:2012tva,Aaij:2013iag,LHCbPprime}. In the effective Hamiltonian approach used to analyse radiative decays at low energies, 
one of the most prominent virtues of this  decay is the capacity to unveil NP contributions inside the short-distance Wilson coefficients, denoted $\C{i}=\C{i}^{\rm SM}+\dC{i}$, not only for the
Standard Model (SM) electromagnetic and dileptonic operators 
\begin{eqnarray}
\ord_7&=&e/(16\pi^2)\, m_b (\bar{s}\sigma_{\mu\nu}P_R b)F^{\mu\nu},\\
\ord_9&=&e^2/(16\pi^2) \, (\bar s\gamma_\mu P_Lb)(\bar \ell \gamma^{\mu} \ell),\\
\ord_{10}&=&e^2/(16\pi^2)\, (\bar s\gamma_\mu P_Lb)(\bar\ell \gamma^{\mu}\gamma_5 \ell),
\end{eqnarray}
(with the usual $P_{L,R}$ chirality projection operators) but also for the chirally-flipped operators $\ord_{i'}$ as well as the scalar and pseudoscalar operators $\ord_{S,P,S',P'}$. Among these, the only non-negligible Wilson coefficients in the SM are $\C{7\rm eff,9,10}^{\rm SM}(\mu_b)=(-0.29,4.07,-4.31)$ at $\mu_b=4.8$~GeV.
The  correlations between the Wilson coefficients constitute  a unique test ground to find consistent patterns pointing towards specific NP models.

However, the presence of hadronic effects can easily hide a NP signal. For this reason, it is essential to design an optimised basis of observables, easy to measure, with low hadronic and high NP sensitivities. In Refs.~\cite{1207.2753,1303.5794} we proposed such a basis, consisting of $P_{1,2,3}$ and $P^\prime_{4,5,6}$ (primary observables with a low sensitivity to form-factor uncertainties at low dilepton invariant mass $q^2$), together with $F_L$ (or $\afb$) and $d\Gamma/dq^2$ (containing large uncertainties but required to complete the basis).

There has been an evolution in the type of observables measured by LHCb. It started with 
the set of observables $\afb$, $F_L$ and $S_3$~\cite{Aaij:2013iag}, all of them rather sensitive to hadronic uncertainties. The experimental results pointed towards a scenario consistent with the SM, but with small deviations in $\afb$ (in both the $q^2$ bin  [2-4.3] GeV$^2$ and the position of the zero).
The next generation of measurements included a theoretically-controlled version of $\afb$ called $A_T^{\rm (re)}$~\cite{becirevic} or $P_2$~\cite{1202.4266}, and $P_1$, which are both less sensitive to hadronic effects and able to magnify deviations due to NP. Finally, LHCb has issued very recent results~\cite{LHCbPprime} completing the basis of $P_i$ and $P_i^\prime$ primary observables~\cite{1202.4266,1207.2753,1303.5794}. These observables, with little sensitivity to hadronic uncertainties at low $q^2$,
have unveiled a set of tensions with respect to the SM that have to be understood from the theoretical point of view. This paper aims at providing such a consistent picture, where the Wilson coefficient $\C{9}$ plays an essential role.

In Sec.~\ref{s1} we discuss the experimental evidence, i.e., the pattern of deviations observed at LHCb.   In Sec.~\ref{sec:2}, we present the main results and the details of our analysis of data using our basis of observables. Finally, in Sec.~\ref{sec:3} we explore the robustness of these results analysing different sources of uncertainty, namely, their sensitivity to perturbative and non-perturbative charm effects, large power-suppressed corrections as well as the comparison between naive and NLO QCD factorisations. We discuss possible improvements on the control of the S-wave pollution in an appendix. We conclude by suggesting cross-checks of our findings and further prospects for similar analyses.

\sect{Experimental evidence}
\label{s1}

The recent LHCb breakthrough, leading to the measurement of most observables of the basis, namely, $P_{1,2}$ and $P_{4,5,6,8}^\prime$ using folded distributions~\cite{Aaij:2013iag,LHCbPprime}, actually exhibits a consistent pattern of deviations with respect to SM expectations. In Table~\ref{TabObs} we summarise the experimental results expressed in our convention in Refs.~\cite{1202.4266,1207.2753}.
Ordering the bins according to the dilepton invariant mass $q^2$, and focusing on the first three bins,
comparing the data~\cite{Aaij:2013iag,LHCbPprime} with the NP scenarios 
discussed in Ref.~\cite{1202.4266} leads to the following comments:

\begin{itemize}
\item There is no substantial deviation in $P_1$ (still with large error bars). From this observable, no definite conclusion can be drawn yet on the presence nor the absence of contributions from right-handed currents. One can notice a mild preference for negative values in the first two bins and positive ones in the third bin, also present in CDF measurements~\cite{1108.0695}. 
\item 
A slight preference for a lower value of the second and third bins of  $\afb$ is consistent with
a 2.9$\,\sigma$ (1.7$\,\sigma$) deviation in the second (third) bin of $P_2$. One notices also
a preference for a slightly higher $q^2$-value for the zero of $\afb$ (at the same position as the zero of $P_2$).
Both effects can be accommodated with $\dC{7}<0$ and/or $\dC{9}<0$.
\item There is a striking $4.0 \,\sigma$ ($1.6 \,\sigma$) deviation in the third (second) bin of $P_5^\prime$~\footnote{In Ref.~\cite{LHCbPprime}, only a $3.7 \,\sigma$ deviation was quoted for the third bin, due to the fact that the comparison was performed between the central experimental value and the 68.3\% CL upper bound for the theoretical prediction. Here the deviation is computed comparing the two central values.}, consistent with large negative contributions in $\dC{7}$ and/or $\dC{9}$. 
\item $P_4^\prime$ is in agreement with the SM, but within large uncertainties, and it has future potential to determine the sign of $\dC{10}$.
\item $P_6^\prime$ and $P_8^\prime$ show small deviations with respect to the SM, but such effect would require complex phases
in $\dC{9}$ and/or $\dC{10}$.
\end{itemize}
A similar pattern of deviations can be observed when one considers the wider $q^2$ bin [1,6] GeV$^2$. Other mechanisms, involving contributions to chirally-flipped operators $\dC{7',9',10'}$, could yield similar effects on some of these observables, but they fail to provide a consistent picture, as will become clear in the following.

Indeed, deviations involving  mainly $P_2$, $P_4^\prime$ and $P_5^\prime$ at low $q^2$ can be understood qualitatively if NP affects $\C{9}$ and/or $\C{10}$. First, at low $q^2$, this NP contribution will impact the transversity amplitudes $A_{\perp,\|}$ only mildly (as it will be hidden by the contribution proportional to $\C{7}$ which is  enhanced by the $1/q^2$ photon pole) but it will affect $A_0$ much more significantly (where $\C{7}$ is not enhanced). At low $q^2$, this type of NP will thus mainly affect observables built from the $A_0$ amplitude such as $P_4^\prime$ and $P_5^\prime$.
Second, as shown in Ref.~\cite{1202.4266}, the position of the zero of $P_2$ and $P_5^\prime$ is affected by a contribution $\dC{7}$ and/or $\dC{9}$, whereas that of $P_4^\prime$ depends on $\dC{7}$, $\dC{9}$ and $\dC{10}$. 
A NP contribution $\dC{9}<0$ would shift the zero
of $P_2$ (resp. $P_4^\prime$ and $P_5^\prime$) to higher (resp. lower and higher) $q^2$ values, leading to an increase of the third bin (resp. an increase of the first bin and an increase of the second and third bins). A contribution from $\dC{10}<0$ would mainly affect the zero of $P_4^\prime$ in a similar way to $\dC{9}<0$, whereas a contribution from $\dC{7}<0$ would have the same impact as $\dC{9}<0$ in $P_2$ and $P_5^\prime$ .


\sect{Analysis, Results and Discussion}
\label{sec:2}

Qualitatively, all the comments of the previous section point towards a scenario where $\dC{9} < 0$ (with possible small contributions $\dC{7}$, $\dC{10} <0$) in a consistent way. We will now proceed with a quantitative analysis of these measurements. 

\begin{table}
\ra{1.12}
\rb{2.5mm}
\begin{tabular}{@{}lrrr@{}}
\toprule[1.2pt]
Observable & Experiment & SM prediction & Pull\\
\midrule[1.1pt]\\[-3.5mm]
 $\av{P_1}_{[0.1,2]}$ & $-0.19^{+0.40}_{-0.35}$ & $0.007^ {+0.043}_{-0.044}$  &  $-0.5$ \\
      $\av{P_1}_{[2,4.3]}$ & $-0.29^{+0.65}_{-0.46}$ & $-0.051^ {+0.046}_{-0.046}$  &  $-0.4$ \\
      $\av{P_1}_{[4.3,8.68]}$ & $0.36^{+0.30}_{-0.31}$ & $-0.117^ {+0.056}_{-0.052}$  &  $+1.5$ \\
      $\av{P_1}_{[1,6]}$ & $0.15^{+0.39}_{-0.41}$ & $-0.055^ {+0.041}_{-0.043}$  &  $+0.5$ \\[0.3mm]
      \midrule[0.5pt]\\[-3.9mm]
      $\av{P_2}_{[0.1,2]}$ & $0.03^{+0.14}_{-0.15}$ & $0.172^ {+0.020}_{-0.021}$  &  $-1.0$ \\
      $\av{P_2}_{[2,4.3]}$ & $0.50^{+0.00}_{-0.07}$ & $0.234^ {+0.060}_{-0.086}$  &  $+2.9$ \\
      $\av{P_2}_{[4.3,8.68]}$ & $-0.25^{+0.07}_{-0.08}$ & $-0.407^ {+0.049}_{-0.037}$  &  $+1.7$ \\
      $\av{P_2}_{[1,6]}$ & $0.33^{+0.11}_{-0.12}$ & $0.084^ {+0.060}_{-0.078}$  &  $+1.8$ \\[0.3mm]
      \midrule[0.5pt]\\[-3.9mm]
      $\av{P_4'}_{[0.1,2]}$ & $0.00^{+0.52}_{-0.52}$ & $-0.342^ {+0.031}_{-0.026}$  &  $+0.7$ \\
      $\av{P_4'}_{[2,4.3]}$ & $0.74^{+0.54}_{-0.60}$ & $0.569^ {+0.073}_{-0.063}$  &  $+0.3$ \\
      $\av{P_4'}_{[4.3,8.68]}$ & $1.18^{+0.26}_{-0.32}$ & $1.003^ {+0.028}_{-0.032}$  &  $+0.6$ \\
      $\av{P_4'}_{[1,6]}$ & $0.58^{+0.32}_{-0.36}$ & $0.555^ {+0.067}_{-0.058}$  &  $+0.1$ \\[0.3mm]
      \midrule[0.5pt]\\[-3.9mm]
      $\av{P_5'}_{[0.1,2]}$ & $0.45^{+0.21}_{-0.24}$ & $0.533^ {+0.033}_{-0.041}$  &  $-0.4$ \\
      $\av{P_5'}_{[2,4.3]}$ & $0.29^{+0.40}_{-0.39}$ & $-0.334^ {+0.097}_{-0.113}$  &  $+1.6$ \\
      $\av{P_5'}_{[4.3,8.68]}$ & $-0.19^{+0.16}_{-0.16}$ & $-0.872^ {+0.053}_{-0.041}$  &  $+4.0$ \\
      $\av{P_5'}_{[1,6]}$ & $0.21^{+0.20}_{-0.21}$ & $-0.349^ {+0.088}_{-0.100}$  &  $+2.5$ \\[0.3mm]
      \midrule[0.5pt]\\[-3.9mm]
      $\av{P_6'}_{[0.1,2]}$ & $0.24^{+0.23}_{-0.20}$ & $-0.084^ {+0.034}_{-0.044}$  &  $+1.6$ \\
      $\av{P_6'}_{[2,4.3]}$ & $-0.15^{+0.38}_{-0.36}$ & $-0.098^ {+0.043}_{-0.056}$  &  $-0.1$ \\
      $\av{P_6'}_{[4.3,8.68]}$ & $0.04^{+0.16}_{-0.16}$ & $-0.027^ {+0.060}_{-0.063}$  &  $+0.4$ \\
      $\av{P_6'}_{[1,6]}$ & $0.18^{+0.21}_{-0.21}$ & $-0.089^ {+0.042}_{-0.052}$  &  $+1.3$ \\[0.3mm]
      \midrule[0.5pt]\\[-3.9mm]
      $\av{P_8'}_{[0.1,2]}$ & $-0.12^{+0.56}_{-0.56}$ & $0.037^ {+0.037}_{-0.030}$  &  $-0.3$ \\
      $\av{P_8'}_{[2,4.3]}$ & $-0.30^{+0.60}_{-0.58}$ & $0.070^ {+0.045}_{-0.034}$  &  $-0.6$ \\
      $\av{P_8'}_{[4.3,8.68]}$ & $0.58^{+0.34}_{-0.38}$ & $0.020^ {+0.054}_{-0.055}$  &  $+1.5$ \\
      $\av{P_8'}_{[1,6]}$ & $0.46^{+0.36}_{-0.38}$ & $0.063^ {+0.042}_{-0.033}$  &  $+1.0$ \\[0.3mm]
      \midrule[0.5pt]\\[-3.9mm]
   $\av{\afb}_{[0.1,2]}$ & $-0.02^{+0.13}_{-0.13}$ & $-0.136^ {+0.051}_{-0.048}$  &  $+0.8$ \\
   $\av{\afb}_{[2,4.3]}$ & $-0.20^{+0.08}_{-0.08}$ & $-0.081^ {+0.055}_{-0.069}$  &  $-1.1$ \\
   $\av{\afb}_{[4.3,8.68]}$ & $0.16^{+0.06}_{-0.05}$ & $0.220^ {+0.138}_{-0.113}$  &  $-0.5$ \\
   $\av{\afb}_{[1,6]}$ & $-0.17^{+0.06}_{-0.06}$ & $-0.035^ {+0.037}_{-0.034}$  &  $-2.0$ \\[0.3mm]
   \midrule[1.1pt]\\[-3.9mm]
      $\av{P_1}_{[14.18,16]}$ & $0.07^ {+0.26}_{-0.28}$ & $-0.352^ {+0.697}_{-0.468}$  &  $+0.6$ \\
      $\av{P_1}_{[16,19]}$ & $-0.71^ {+0.36}_{-0.26}$ & $-0.603^ {+0.589}_{-0.315}$  &  $-0.2$ \\[0.3mm]
      \midrule[0.5pt]\\[-3.9mm]
      $\av{P_2}_{[14.18,16]}$ & $-0.50^ {+0.03}_{-0.00}$ & $-0.449^ {+0.136}_{-0.041}$  &  $-1.1$ \\
      $\av{P_2}_{[16,19]}$ & $-0.32^ {+0.08}_{-0.08}$ & $-0.374^ {+0.151}_{-0.126}$  &  $+0.3$ \\[0.3mm]
      \midrule[0.5pt]\\[-3.9mm]
      $\av{P_4'}_{[14.18,16]}$ & $-0.18^ {+0.54}_{-0.70}$ & $1.161^ {+0.190}_{-0.332}$  &  $-2.1$ \\
      $\av{P_4'}_{[16,19]}$ & $0.70^ {+0.44}_{-0.52}$ & $1.263^ {+0.119}_{-0.248}$  &  $-1.1$ \\[0.3mm]
      \midrule[0.5pt]\\[-3.9mm]
      $\av{P_5'}_{[14.18,16]}$ & $-0.79^ {+0.27}_{-0.22}$ & $-0.779^ {+0.328}_{-0.363}$  &  $+0.0$ \\
      $\av{P_5'}_{[16,19]}$ & $-0.60^ {+0.21}_{-0.18}$ & $-0.601^ {+0.282}_{-0.367}$  &  $+0.0$ \\[0.3mm]
      \midrule[0.5pt]\\[-3.9mm]
      $\av{P_6'}_{[14.18,16]}$ & $0.18^ {+0.24}_{-0.25}$ & $0.000^ {+0.000}_{-0.000}$  &  $+0.7$ \\
      $\av{P_6'}_{[16,19]}$ & $-0.31^ {+0.38}_{-0.39}$ & $0.000^ {+0.000}_{-0.000}$  &  $-0.8$ \\[0.3mm]
      \midrule[0.5pt]\\[-3.9mm]
      $\av{P_8'}_{[14.18,16]}$ & $-0.40^ {+0.60}_{-0.50}$ & $-0.015^ {+0.009}_{-0.013}$  &  $-0.6$ \\
      $\av{P_8'}_{[16,19]}$ & $0.12^ {+0.52}_{-0.54}$ & $-0.008^ {+0.005}_{-0.007}$  &  $+0.2$ \\[0.3mm]
      \midrule[0.5pt]\\[-3.9mm]
   $\av{\afb}_{[14.18,16]}$ & $0.51^ {+0.07}_{-0.05}$ & $0.404^ {+0.199}_{-0.191}$  &  $+0.5$ \\
   $\av{\afb}_{[16,19]}$ & $0.30^ {+0.08}_{-0.08}$ & $0.360^ {+0.205}_{-0.172}$  &  $-0.3$ \\[0.3mm]
\midrule[1.1pt]\\[-3.8mm]
$10^4\, {\cal B}_{B\to X_s\gamma} $ & $3.43\pm 0.22$ & $3.15\pm 0.23$ & $+0.9$\\
$10^6\, {\cal B}_{B\to X_s\mu^+\mu^-}$ & $1.60\pm 0.50$ & $1.59\pm 0.11$ & $+0.0$\\
$10^9\, {\cal B}_{B_s\to \mu^+\mu^-} $ & $2.9\pm 0.8$ & $3.56\pm 0.18$ & $-0.8$\\
$A_I(B\to K^* \gamma)$ & $0.052\pm 0.026$ & $0.041\pm 0.025$ & $+0.3$\\
$S_{K^* \gamma}$ & $-0.16\pm 0.22$ & $-0.03\pm 0.01$ & $-0.6$\\[0.2mm]
\bottomrule[1.2pt]
\end{tabular}
\caption{Experimental averages and SM predictions for the observables used in the analysis. The dictionary between the different conventions used in Refs.~\cite{Aaij:2013iag,LHCbPprime} and \cite{1202.4266,1207.2753} is: $P_1^{\textrm{LHCb}}=P_1$, $P_2^{\textrm{LHCb}}=-P_2$, $P_4^{\prime {\rm LHCb}}=-\frac{1}{2} P_4^\prime$, $P_5^{\prime {\rm LHCb}}=P_5^\prime$, $P_6^{\prime {\rm LHCb}}=-P_6^\prime$ and $P_8^{\prime \rm LHCb}=\frac{1}{2} P_8^\prime$ (sign differences stem from $\theta_\ell^{\textrm{LHCb}}=\pi-\theta_\ell$).
\label{TabObs}}
\end{table}

\subsection{General analysis and overview of constraints}\label{sec:overview}

We start with a global analysis to the data, in a general scenario with simultaneous arbitrary (real) NP contributions to the Wilson coefficients  $\C{7,9,10}$ and $\C{7',9',10'}$, writing $\C{i}= \C{i}^{\rm SM}+\dC{i}$
 (we neglect scalar contributions).  We use the predictions and uncertainty estimates described in Ref.~\cite{1303.5794}, following NLO QCD factorisation for the large-recoil (low-$q^2$) bins~\cite{0106067,0412400} and HQET for the low-recoil (large-$q^2$) bins~\cite{grinstein+pirjol}. We follow a standard $\chi^2$  frequentist approach, following App.~C in Ref.~\cite{1207.2753}: we take into account the asymmetric errors on the experimental numbers, estimate theoretical uncertainties for each set of values for $\dC{i}$, and treat all uncertainties (experimental and theoretical) as statistical to combine them in quadrature. The correlations among the measurements are not available currently and are thus neglected. We consider the following observables:\\[-2mm]

\nn 1. \emph{Optimised observables in $B\to K^*\mu^+\mu^-$}: We take $P_{1}$, $P_2$, $P_{4}^\prime$, $P_5^\prime$, $P_6^\prime$ and $P_8^\prime$, within the 3 large-recoil bins [0.1,2], [2,4.3] and [4.3,8.68] GeV$^2$, and the 2 low-recoil bins [14.18,16] and [16,19]  GeV$^2$. We note that: \emph{a)} all these observables are independent, as $P_3$ is not measured (see Ref.~\cite{1202.4266}), \emph{b)} these observables have little hadronic sensitivity only at low $q^2$ \cite{1303.5794}, so we expect weak constraints from the low-recoil region, \emph{c)} we do not consider the [1,6] bin at this stage since these observables are not independent of the previous ones. We will consider this bin later on.\\[-2mm]

\nn 2. \emph{Forward-Backward Asymmetry in $B\to K^*\mu^+\mu^-$}: Once one has chosen a maximal set of optimised observables, one has still to choose two independent observables sensitive to form-factor uncertainties. The differential branching ratio $d{\cal B}/dq^2$ is one of them, necessary to fix the overall normalisation. We do not include this observable because of its large theoretical uncertainty derived from its significant sensitivity to hadronic form factors. The other observable can be either $\afb$ or $F_L$. We choose $\afb$ because of its expected higher sensitivity to $\dC9$ and its complementarity with $P_2$~\cite{1202.4266,matias-Eps}. Again, we keep the [1,6] bin for a later stage of the analysis. We consider only LHCb measurements (the inclusion of the available results from other experiments has only a marginal impact on the data~\cite{hfag}).\\[-2mm]

\nn 3. \emph{Radiative and dileptonic $B$ decays}: There are other important $b\to s$ penguin modes sensitive to magnetic and dileptonic operators. We consider the branching ratios ${\cal B}(B\to X_s\gamma) _{E_\gamma>1.6 {\rm GeV}}$, ${\cal B}(B\to X_s \mu^+\mu^-)_{[1,6 ]}$ and ${\cal B}(B_s\to \mu^+\mu^-)$, the isospin asymmetry $A_I(B\to K^* \gamma)$ and the  $B\to K^* \gamma$ time-dependent CP asymmetry $S_{K^*\gamma}$. Relevant formulas for these observables can be found in Ref.~\cite{1104.3342}, while updated experimental numbers are taken from Refs.~\cite{hfag,0712.3009,1211.2674}
and Refs.~\cite{LHCbBsmumu-Eps,CMSBsmumu-Eps,FlavourExp-Eps} (where the average for ${\cal B}(B_s\to \mu^+\mu^-)$ takes into account differences in the ratio of production fractions $f_s/f_d$ and normalisations for CMS and LHCb). We disregard other similar observables, either because their theoretical description is not ascertained, such as $A_{CP}(B\to X_s\gamma)$, or because of experimental issues, as is the case with $B\to K\mu^+\mu^-$ due to the unclear status of the experimental separation of neutral and charge modes indicated by the measured
isospin asymmetry~\cite{1205.3422}. 

For $B\to K\mu^+\mu^-$, an additional issue was raised in Ref.~\cite{Aaij:2013pta}, as an unexpected resonant structure $\psi(4160)$ has been observed in $B^+\to K^+ \mu^+ \mu^-$ at low recoil. It remains to be seen how this resonant structure can impact the neutral mode around $q^2\simeq 17.3$ GeV$^2$, and if it can modify the predictions for $B\to K^* \mu^+ \mu^-$ observables for the two bins in the low-recoil region. In the following analysis, we will always consider two data sets: one with only large-recoil data, the other one with both low- and large-recoil data.

Experimental averages and SM theoretical predictions for all these observables are summarised in Table~\ref{TabObs}. \mbox{As an} outcome of the fit, all Wilson coefficients have $2\,\sigma$ C.L. intervals encompassing the SM value, except for $\C9$ (below its SM value) with a best fit point around $\dC9 \sim -1.2$. The SM hypothesis ($\dC{i}=0$ for all $i$) has a pull\footnote{When testing a hypothesis (e.g. the SM) in a given framework, we refer to its p-value in units of $ \sigma$ as the ``pull" of this hypothesis. Therefore, a lower p-value corresponds to a larger pull in units of $\sigma$.}  of 2.9$\,\sigma$. 
The individual 1, 2 and 3$\,\sigma$ intervals for the Wilson coefficients are given in Table~\ref{tab1}~\footnote{Besides the region close to the SM point, there is an allowed but less favoured region close to the flipped-sign solution for the Wilson coefficients ($\C{i}\simeq -\C{i}^{\rm SM}$). The appearance of this region is expected and well understood (e.g., Ref.~\cite{1104.3342}), but would correspond to a significant amount of NP. We thus disregard it for the time being, even though this should be kept in mind in future analyses.}.
The most economical scenario corresponds to a negative NP contribution to $\C9$ with all the other Wilson coefficients close to their SM value. Even though the branching ratio is affected by very large uncertainties and is not considered in our analysis, it is also interesting to notice that $\dC9<0$ would tend to decrease the differential branching ratio, improving the agreement with experimental data.

\begin{table}
\ra{1.15}
\rb{2.5mm}
\begin{tabular}{@{}lccc@{}}
\toprule[1.2pt]
Coefficient & 1$\,\sigma$ & 2$\,\sigma$ & 3$\,\sigma$ \\
\midrule[1.1pt]\\[-3.5mm]
$\dC7$      &  $[-0.05,-0.01]$  &  $[-0.06,0.01]$  &  $[-0.08,0.03]$ \\ 
$\dC9$      &  $[-1.6,-0.9]$  &  $[-1.8,-0.6]$  &  $[-2.1,-0.2]$ \\ 
$\dC{10}$   &  $[-0.4,1.0]$  &  $[-1.2,2.0]$  &  $[-2.0,3.0]$ \\ 
$\dCp7$     &  $[-0.04,0.02]$  &  $[-0.09,0.06]$  &  $[-0.14,0.10]$ \\ 
$\dCp9$     &  $[-0.2,0.8]$  &  $[-0.8,1.4]$  &  $[-1.2,1.8]$ \\ 
$\dCp{10}$  &  $[-0.4,0.4]$  &  $[-1.0,0.8]$  &  $[-1.4,1.2]$ \\
\bottomrule[1.2pt]
\end{tabular}
\caption{$68.3\%$ (1$\,\sigma$), $95.5\%$ (2$\,\sigma$) and $99.7\%$ (3$\,\sigma$) confidence intervals for the NP contributions to Wilson coefficients resulting from the global analysis.
\label{tab1}}
\end{table}

The large-recoil data favours $\dC9<0$ more significantly than the low-recoil region. Removing the low-recoil $B\to K^*\mu^+\mu^-$ observables from the fit enhances the effect, with a best fit point around $\dC9\sim -1.6$. In this case also a negative contribution $\dCp9<0$ is favoured. This pattern is also obtained when considering only the $[1,6]$ bin, though with larger uncertainties. We discuss these issues in more depth below. 
  
\subsection{Comparison of NP scenarios}

In order to clarify the role played by $\C9$ in explaining the $B\to K^*\mu^+\mu^-$ anomaly and to discuss the role of the other coefficients, it is interesting to consider nested scenarios where NP is turned on for each Wilson coefficient one after the other, starting from the SM hypothesis. In a given scenario (where some Wilson coefficients $\C{j_1,\ldots j_N}$ receive NP and the others do not), the
improvement obtained by allowing one more Wilson coefficient $\C{i}$ to receive NP contributions can be quantified by computing the pull of the 
$\dC{i}=0$ hypothesis within the scenario where $\dC{i}$ and  $\dC{j_1,\ldots j_N}$ are left free~\cite{Lenz:2010gu}.

When considering the full set of large- and low-recoil data for $B\to K^*\mu^+\mu^-$, we find that the larger pulls (around $\sim 4\,\sigma$) are obtained when adding $\dC9$, independently of which other Wilson coefficients are left free to receive NP contributions. The next-to-larger pulls are obtained by adding $\dC7$ (around $\sim 3\,\sigma$), in all cases except when $\dC9$ has been added beforehand; in such a case, the pull is $\sim 1.3\,\sigma$ (still the lowest after $\dC9$). The rest of the pulls are always around or below $1\,\sigma$. These results are consistent with the fact that  $\C9$ plays a prominent role in explaining the $B\to K^*\mu^+\mu^-$ anomaly; besides that, a NP contribution to $\C7$ is also favoured
though less strongly.

It is also interesting to consider only the large-recoil bins, for which the theoretical description of the optimised observables is more accurate. The main picture remains the same, although in some cases some other coefficients may play a (more modest) role in the discussion. For instance, pulls around $\sim 3\,\sigma$ can be obtained for $\C{10}$ and $\Cp{10}$ if we insist that $\dC9=0$ (not otherwise). Another example arises in a scenario with only $\dC9\ne 0$, for which the next most relevant NP contributions are $\C7$ and $\Cp9$ (with pulls at the same level).

Finally, using both low- and large-recoil data  for \mbox{$B\to K^*\mu^+\mu^-$}, we can compute the pull 
corresponding to the $\dC{10,7',9',10'}=0$ hypothesis in the scenario where all 6 Wilson coefficients are allowed to receive NP contributions. The resulting pull is below $\sim 0.2\,\sigma$, indicating that no other NP contribution is required by the data apart from $\dC9$ and $\dC7$. The same results occur when only large-recoil data is used.

Before focusing on this scenario in the following, a comment is in order concerning alternative scenarios with different sets of coefficients  receiving NP contributions. In all scenarios considered the best fit corresponds to $\dC9\sim -1.2$ with a significant preference for negative values. In addition, a slight preference for negative values of $\dCp9$ or $\dC7$ occurs  (with much less significance). It arises for $\dCp9$ when only large-recoil data is considered: $\dCp9<0$ is favoured to raise the value of $\av{P_5^\prime}_{[4.3,8.68]}$ without spoiling the agreement between theory and experiment in the first bin. This possibility is however weakened by the low-recoil data, and we will not consider this possibility any further.
On the other hand, $\dC7<0$ is also favoured by the radiative decays (see e.g. Ref~\cite{1207.2753}), and deserves further consideration.

\subsection{$(\C7^{\rm NP},\C9^{\rm NP})$ scenario}\label{sec:C7C9}

\begin{figure}
\includegraphics[width=8.5cm]{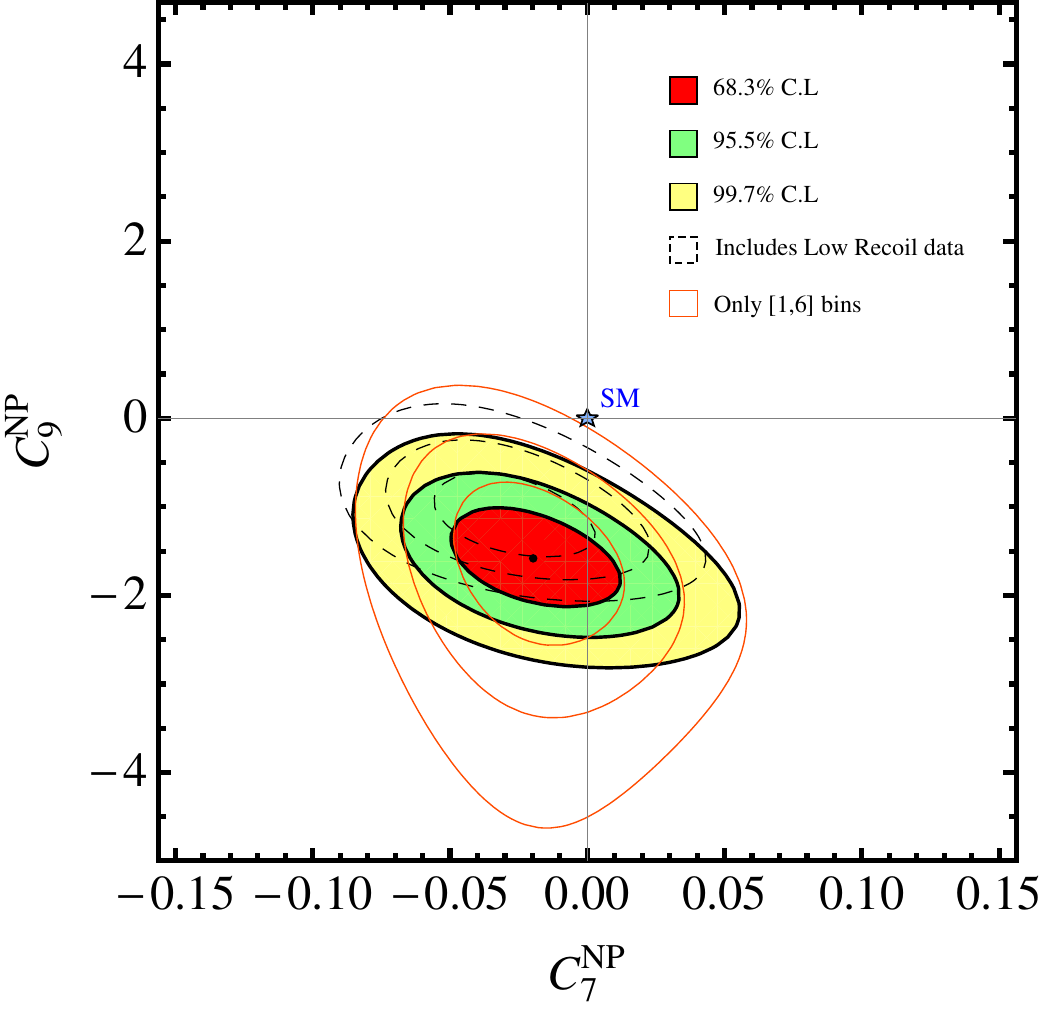}
\caption{Fit to $(\dC7,\dC9)$, using the three large-recoil bins for $B\to K^*\mu^+\mu^-$ observables, together with $B\to X_s\gamma$, $B\to X_s\mu^+\mu^-$, $B\to K^*\gamma$ and $B_s\to \mu^+\mu^-$. The dashed contours include both large- and low-recoil bins, whereas the orange (solid) ones use only the 1-6 GeV$^2$ bin for $B\to K^*\mu^+\mu^-$ observables. The origin $\dC{7,9}=(0,0)$ corresponds to the SM values for the Wilson coefficients $\C{7{\rm eff},9}^{\rm SM}=(-0.29,4.07)$ at $\mu_b=4.8$ GeV.}
\label{fig}
\end{figure}

\begin{figure*}
\includegraphics[width=5.5cm]{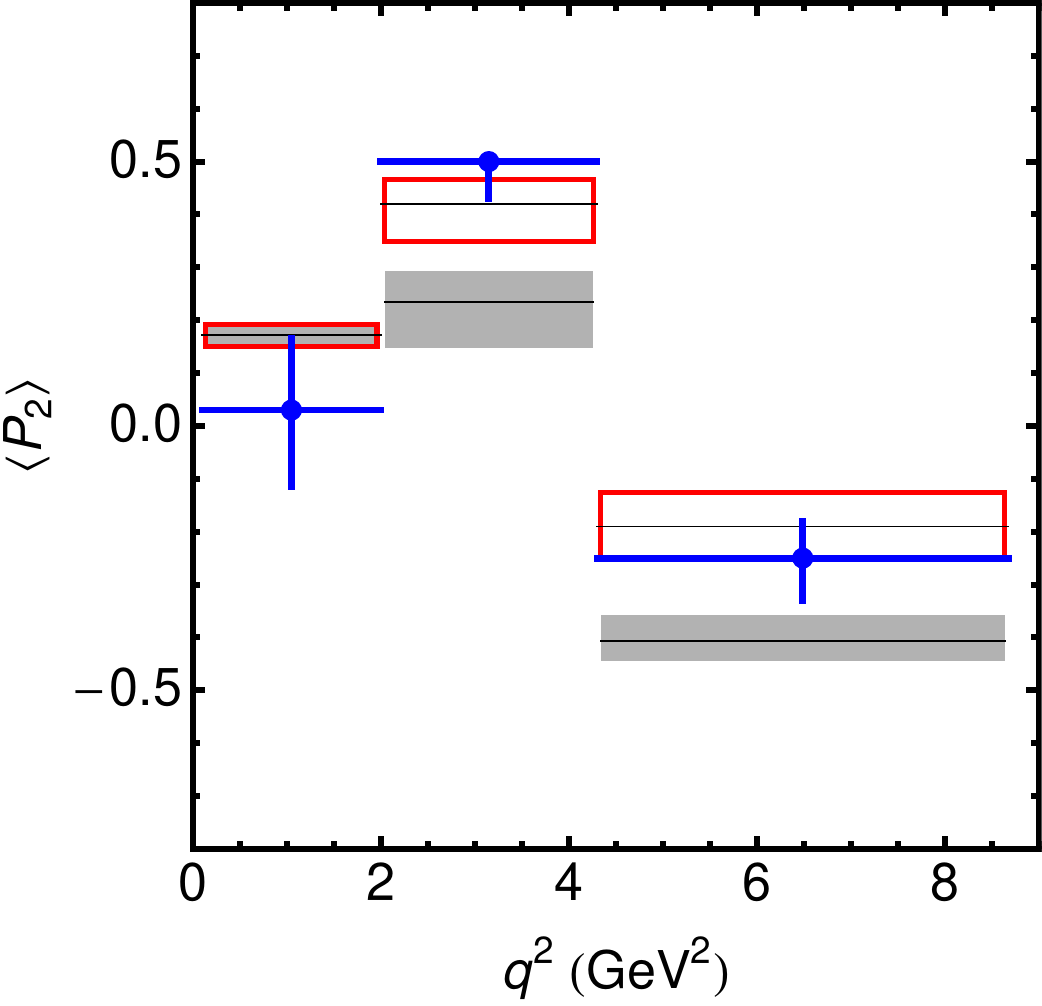}\hspace{0.4cm}
\includegraphics[width=5.5cm]{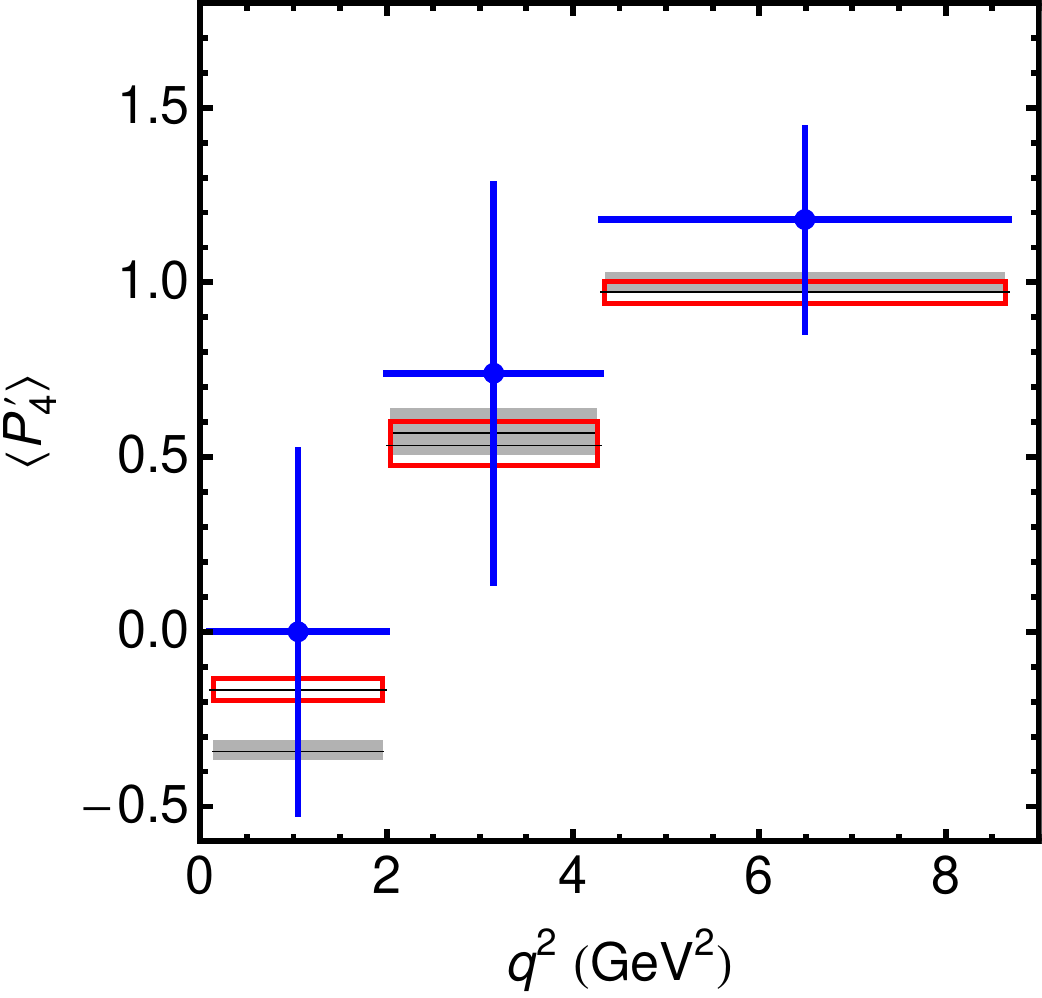}\hspace{0.4cm}
\includegraphics[width=5.5cm]{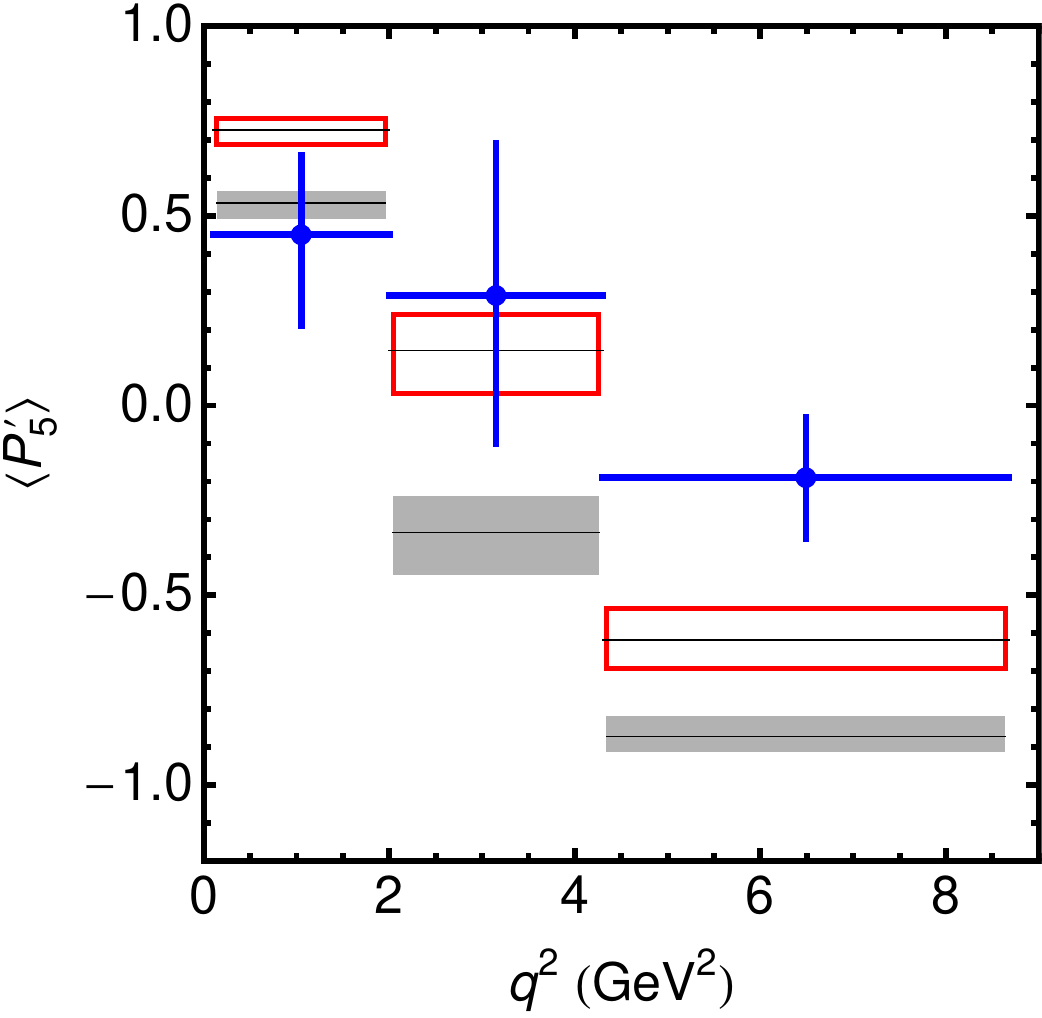}\\[5mm]
\includegraphics[width=5.5cm]{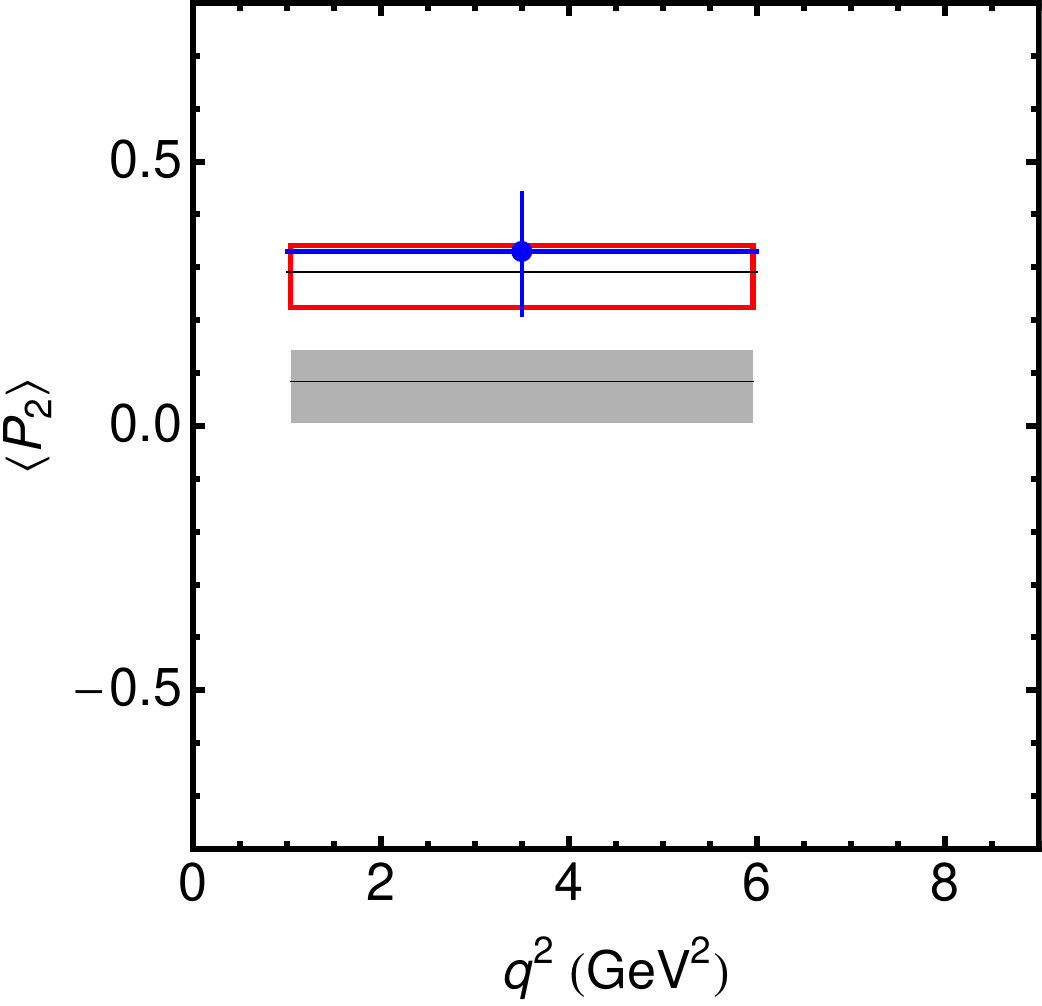}\hspace{1cm}
\includegraphics[width=5.5cm]{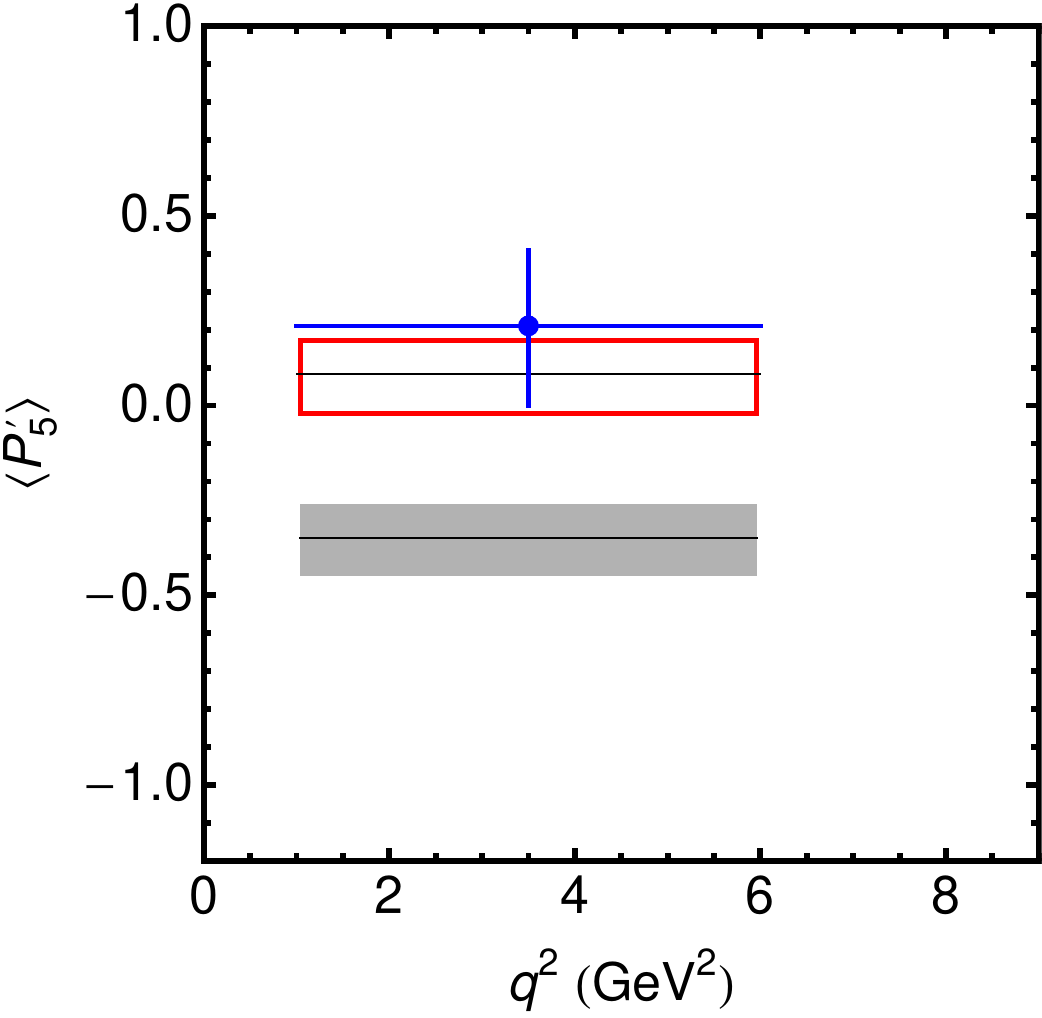}
\caption{Comparison between the SM predictions (gray boxes), the experimental measurements (blue data points) and the predictions for the scenario with $\dC9=-1.5$ and other $\dC{i}=0$ (red squares).}
\label{fig2}
\end{figure*}
We focus on the implications of data for NP in $\C7$ and $\C9$. We perform a standard $\chi^2$ fit to $\dC7$, $\dC9$, first including all the observables considered in Sec.~\ref{sec:overview}, but taking only the first three large-recoil bins for $B\to K^*\mu^+\mu^-$. The result is shown in Fig.~\ref{fig}, where 68.3\% (red), 95.5\% (green), and 99.7\% (yellow) C.L. regions are shown. The best fit point is obtained for $\dC9\sim -1.5$ and $\dC7\sim -0.02$. We stress that in this scenario, the SM hypothesis $\dC7=0,\dC9=0$ has a pull of 4.5$\,\sigma$. 

The individual $68.3\% {\rm C.L.}$ (1$\,\sigma$) ranges for $\dC7$  and $\dC9$ are
\eq{
\dC7=[-0.035,0.000]\,, \ \ \dC9=[-1.9,-1.3] \,.
}

Including the low-recoil bins decreases slightly the significance of the effect, since these observables are less constraining and 
compatible with the SM currently. The corresponding regions are indicated by the dashed curves in Fig.~\ref{fig}. The inclusion of the low-recoil data reduces the discrepancy with respect to the SM to $3.9\,\sigma$.
We notice that if the (first low-recoil) [14.18,16] bin is excluded from the analysis, we get similar significances analysing either large-recoil data alone or large- and low-recoil data together, meaning that the reduction of significance stems mainly from the [14.18,16] bin, a region where the role of the newly-found $\psi(4160)$ resonance has still to be understood~\cite{LHCbPprime}. 
In both analyses (with or without low-recoil data), $P_2$ and $P_5'$ drive the fits away from the SM point, and fits including only one of these two observables for $B\to K^*\mu^+\mu^-$, together with the other radiative and dileptonic decays, lead to contours in the $(\dC7,\dC9)$ plane similar to Fig.~\ref{fig}. 

We would like to understand whether this conclusion is due to peculiarities of individual bins. For this purpose we repeat the analysis restricting the input for the $B\to K^*\mu^+\mu^-$ observables to $[1,6]$ GeV$^2$ bins,
exploiting several theoretical and experimental advantages.
Such wider bins collect more events with larger statistics. Furthermore, some theoretical issues are less acute, such as the effect of low-mass resonances at very low $q^2\lesssim 1$ GeV$^2$~\cite{camalich}, or the impact of charm loops above $\sim 6$ GeV$^2$~\cite{1006.4945}.  On the other hand, integrating over such a large bin washes out some effects related to the $q^2$ dependence of the observables, so that we expect this analysis to have less sensitivity to NP~\cite{1207.2753}. This can be seen in Fig.~\ref{fig}, where the regions in this case are indicated by the orange curves, and as expected the constraints get slightly weaker. In addition, due to the fact that theoretical uncertainties happen to increase moderately for large negative NP contributions to $\C9$, the constraints are looser in the lower region of the $(\dC7,\dC9)$ plane.  We emphasise that even in this rather conservative situation the main conclusion (a NP contribution $\dC9\sim -1.5$) still prevails, whereas the SM hypothesis has still a pull of 3.2$\,\sigma$. 

We illustrate the improvement gained by shifting $\C{9}$ in Fig.~\ref{fig2}, where we show the predictions for  $\dC9=-1.5$ (and other $\dC{i}= 0$) for the observables $P_2$, $P_4'$ and $P_5'$, together with the experimental data and  SM predictions. In particular, we observe how the various observables described in Sec.~\ref{s1} change for $\dC{9}<0$. If the data is in general well reproduced in this scenario, there are still a few observables difficult to explain theoretically. Looking at Fig.~\ref{fig2}, the most obvious cases are $\av{P_5'}$ in the first and third bins. One can see there is a tension between these two bins: more negative values for $\dC9$ reproduce better the third bin, but drive the first bin upwards, whose experimental value is consistent with the SM. A similar situation happens with the second and third bins of $\av{P_2}$, although in this case a good compromise is achieved. 

Concerning the individual constraints to the fit,
the large-recoil bins for $P_2$ and $P_5^\prime$ both favour the same large region away from the SM in the $(\dC7,\dC9)$ plane, providing a negative correlation between $\dC7$ and $\dC9$. $B\to X_s\gamma$ selects values of $\dC7$ close to the SM value, leading to the combined (smaller) region shown in Fig.~\ref{fig}. To be more quantitative,
we have considered the pulls obtained by removing in turn one or two observables from the fit. We find that the largest pulls are associated to $\av{P_5'}_{[4.3,8.68]}$,  $B\to X_s\gamma$, $\av{P_2}_{[14.18,16]}$ and $\av{P_4'}_{[14.18,16]}$. $B\to X_s\gamma$ has a large pull because it plays a very important role in disfavouring a scenario with large and negative $\dC7$, which can mimic the $\dC9$ scenario in $B\to K^* \mu^+\mu^-$ observables. 
The observables $\av{P_5'}_{[4.3,8.68]}$ and $\av{P_2}_{[14.18,16]}$ pull in different directions: the former favours more negative and the latter less negative values for $\dC9$, while the best fit point lies somewhat in the middle, with or without these observables. On the other hand $\av{P_4'}_{[14.18,16]}$ has a marginal effect on the results of the fit.

The role of individual observables is confirmed by comparing our analysis with the preliminary results in Ref.~\cite{matias-Eps}, performed in the same framework, but with only $P_1$,$P_2$ and $\afb$ as inputs for $B\to K^*\mu^+\mu^-$, leading to a $3\, \sigma$ deviation from the SM in the $(\dC7,\dC9)$ plane (in our present analysis, this effect is magnified by the addition of $P_{4,5,6,8}'$~\cite{LHCbPprime} among the observables). We emphasise the importance of choosing the right set of observables among the three correlated inputs $\afb, P_2,F_L$:  
$F_L$ has a very significant dependence on the choice of form factors (Fig.~\ref{figsm}), which is less acute in the case of $\afb$ and $P_2$, so that the choices $(F_L, P_2)$ or $(F_L, \afb)$ \cite{vandyk-Eps}   lead to results that are more biased by the specific parametrisation of form factors considered and less sensitive to NP compared to $(\afb,P_2)$ \cite{matias-Eps}. For this reason, we use $\afb$ instead of $F_L$ in our analysis. We have checked by two different procedures (NLO QCD factorisation and naive factorisation) that the $3\, \sigma$ deviation reported in Ref.~\cite{matias-Eps}  using [1-6] bins gets reduced to around 1 $\sigma$  if $F_L$ is used as an input instead of $P_2$ or $\afb$ (in agreement with Ref.~\cite{vandyk-Eps}, where $F_L$ is used).

\sect{Robustness of the results}\label{sec:3}

In view of the results of the previous section, it is important to assess the robustness of the NP interpretation for the $B\to K^*\mu^+\mu^-$ anomaly and how stable the conclusion $\dC9<0$ is, taking into account potential pollution from SM sources mimicking a negative $\dC9$. 

\subsection{Charm Loop}

One of the key sources of uncertainty in the extraction of $\C9$ from $B\to K^*\mu^+\mu^-$ is related to the charm-loop contribution (subsequently decaying through a photon into a dilepton pair) coming from the insertion of 4-quark current-current ($\ord_{1,2}^c$) or penguin  operators ($\ord_{3-6}$).  The contributions from $\ord_{1,2}^c$ are particularly important since the Wilson coefficients are numerically large and the processes are not CKM suppressed. This contribution can be described through a short-distance (perturbative) contribution, which exhibits a noticeable sensitivity to the value of $m_c$ near the threshold of $c\bar c$ production, and a long-distance (non-perturbative) contribution which is difficult to assess.

The perturbative charm-loop contribution is usually absorbed into the definition of $\C9^{\rm eff}(q^2)=\C9+Y(q^2)$ \cite{0106067} and is given at leading order by
\eqa{ \label{eq:Y}
Y^c(q^2,m_c) &=& -\frac4{27} (4 \C1+3\C2+18\C3+180\C5)\times \\
&& \hspace{-2.1cm} \times \bigg[ \ln \frac{m_c^2}{\mu^2} -\frac23 -z +(2+z) \sqrt{|z-1|} \,\arccot \sqrt{(z-1)} \bigg] \nb
}
where $z=4 m_c^2/q^2$. There is a threshold at $q^2=4m_c^2\simeq 6$~GeV$^2$, above which
Eq.~(\ref{eq:Y}) must be continued analytically and an imaginary part is generated. The real part exhibits a cusp at this threshold, whose exact position depends on $m_c$. There is a significant variety of choices in the literature concerning the value of $m_c$ for such computation, for instance the pole mass (around 1.4 GeV)~\cite{0106067}, the $\overline{\rm MS}$ mass at the scale $\mu=m_c$ (around 1.27 GeV)~\cite{0512066} or the same mass at the scale $\mu=2m_c$ (around 1 GeV)~\cite{1006.4945}.
Following Ref.~\cite{1303.5794}, we take the second option and perform the computation of $B\to K^*\mu^+\mu^-$ observables with a reference value $\overline m_c=1.27$ GeV. We can study the dependence on $m_c$ by reinterpreting its change as a shift in the value of $\C9$, given by:
\eq{
\delC9^{c\bar c, {\rm pert}} = {\rm Re}[Y^c(q^2,m_c)-Y^c(q^2,\overline m_c)]\ .
\label{ccpert}}
The same analysis can be performed for the imaginary part, which vanishes below the charm threshold, with similar conclusions.

\begin{figure}
\includegraphics[width=8.5cm]{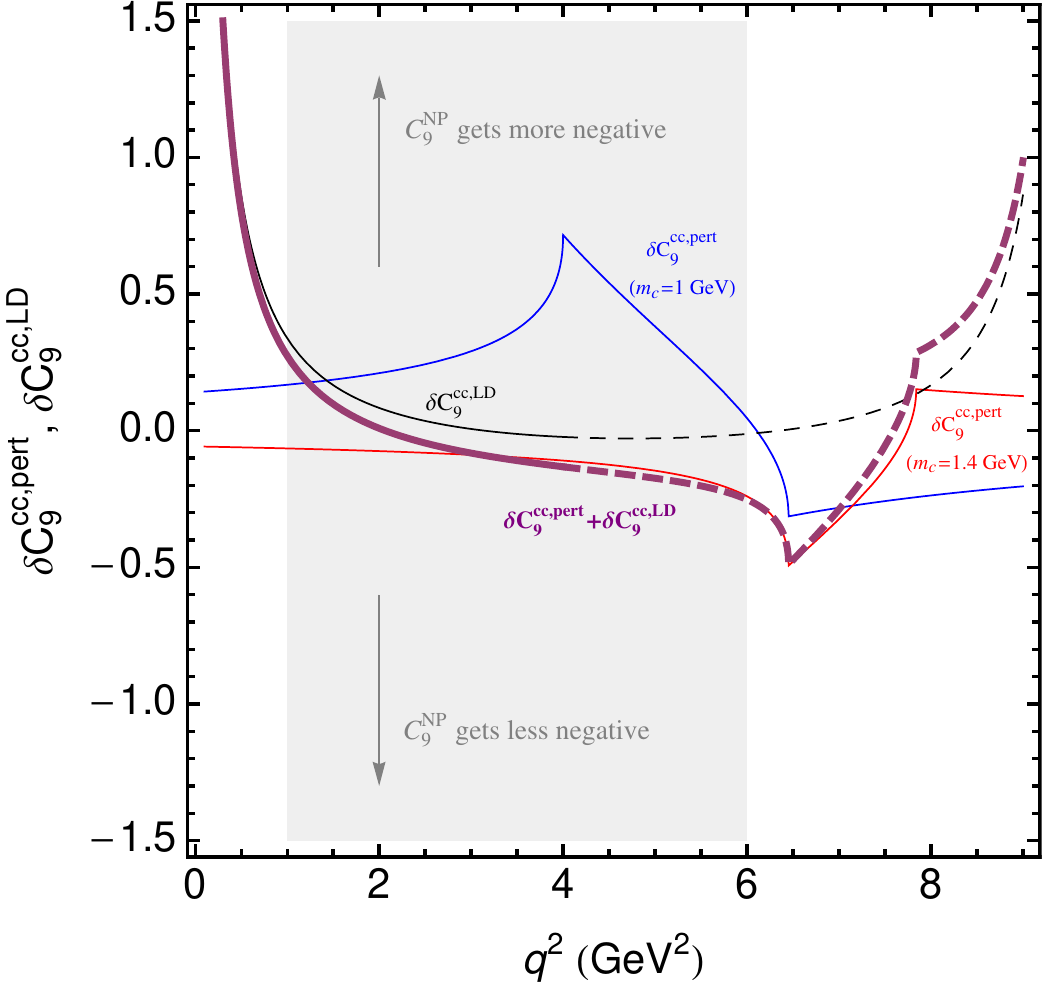}
\caption{Estimate of possible effective corrections to $\C9$ from the $m_c$-dependence of short-distance ($\delC9^{c\bar c, {\rm pert}} $) and long-distance charm-loop effects ($\delC9^{c\bar c, {\rm LD}} $). The thick line shows a (worst-case scenario) combination of both effects. A positive contribution here enhances a negative NP contribution to $\C9$, while a negative contribution here means $\dC9$ is less negative by the same amount. Note that these effects cannot mimic a NP contribution as large as  $\dC9\sim -1.5$. }
\label{figcharm}
\end{figure}

The long-distance contribution is difficult to estimate, and we are not aware of a systematic approach able to compute this correction from first principles. Part of this effect is taken into account directly at the level of the experimental analysis, through the removal of charmonium contributions. From the theoretical side, one can
exploit the light-cone sum rule computation in Ref.~\cite{1006.4945}, performed below the charm threshold and extrapolated up to the $J/\psi$ pole at $q^2 \sim 9.4\ {\rm GeV}^2$. These long-distance contributions
 can be recast as an effective contribution to $\C9$, but they depend on the transversity amplitude considered
 (contrary to the perturbative case). Even though one could in principle distinguish this effect from a ``universal'' NP contribution $\dC9$ by comparing observables with different sensitivities to the three transversity amplitudes, this cannot be achieved with the current uncertainties. In particular, the  results quoted in Ref.~\cite{1006.4945} for each amplitude are compatible within errors. Therefore we consider a universal correction to $\C9$  arising from the long-distance charm-loop contribution, that we parametrize as:
\eq{
\delC9^{c\bar c, {\rm LD}} =\frac{a + b\,q^2 (c-q^2)}{q^2 (c -q^2)}\ .
\label{ccLD}}
The parameters $a,b,c$ can be obtained by imposing that the sum of the perturbative contribution $Y^c$ and the non-perturbative $\delC9^{c\bar c, {\rm LD}}$ contributions reproduce the results in Ref.~\cite{1006.4945} below the charm threshold (correcting for the different reference value of the charm quark mass $\bar m_c(m_c)=1.27$~GeV used here). The parameter $c$ is chosen to recover the $J/\psi$ pole, leading to $a\sim 5$, $b\sim 0.25$ and $c\sim 9.5$.

In Fig.~\ref{figcharm} we show the size and $q^2$ dependence of the perturbative and non-perturbative charm-loop contributions in Eqs.~(\ref{ccpert}) and (\ref{ccLD}). 
The results of Sec.~\ref{sec:2} were obtained taking $Y^c(\bar m_c)$ and neglecting non-perturbative contributions.
To keep a good agreement with data,
an additional $\delta \C9^{c\bar c}$ correction must be compensated by a shift in the NP contribution $\dC9$ compared to the values obtained previously.
A positive $\delta \C9^{c\bar c}$ means that $\dC9$ should be enhanced (more negative), whereas a negative $\delta \C9^{c\bar c}$ requires $\dC9$ to be reduced in magnitude (less negative).
Several comments are in order. First, one can see that the region between 1 and 6 GeV$^2$ is little affected by the long-distance contribution, and gets either an enhancement of NP effects in $\dC9$ (for $m_c=1$ GeV) or a very small depletion (for $m_c=1.4$ GeV) from the short-distance part. In the case of the [4.3,8.68] bin, one has to remember that the negative contribution $\delta \C9^{c\bar c,{\rm pert}}$ when varying $m_c$ from 1.27 to 1.4 GeV has to be integrated over the whole bin, leading to a shift of at most $+0.3$ for $\dC9$ from this bin
(one finds a contribution of similar size for the imaginary part of $\C9$, which is however more difficult to interpret directly in terms of a shift of $\dC9$, taken real here).

We see therefore that \emph{a)} charm-loop effects affect very marginally our analysis using $B\to K^*\mu^+\mu^-$ observables in the [1,6] bin only, and \emph{b}) these long-distance charm contributions will tend to enhance (rather than reduce) the need for a negative $\dC9$ for the analysis including thinner bins, closer to the photon and $J/\psi$ poles.

We have checked the sensitivity of our analysis to these effects (charm-loop effects and charm-mass dependence) by considering the least favourable situation
for NP, where the charm quark mass is shifted from 1.27 GeV to 1.4 GeV and long-distance contributions are neglected.
The need for $\dC9\neq 0$ remains, with a slight decrease in significance (by less than $1\,\sigma$) and a small shift of the best-fit value for $\dC9$ (by $\simeq +0.3$, in agreement with our previous discussion).

\subsection{Alternative approaches to factorisation}

A second issue consists in the sensitivity to the framework used to describe $B\to K^*\mu^+\mu^-$ form factors and amplitudes. Following Refs.~\cite{0106067,1202.4266,1303.5794}, we have adopted the QCD factorisation framework.
One issue still under discussion is the uncertainty coming from $\Lambda/m_b$ corrections. 
As a check of the robustness of our results with respect to this issue, we have increased the uncertainties coming from 
$\Lambda/m_b$ corrections by 3, resulting in a slight decrease of the pull for the SM hypothesis in the $(\C7,\C9)$ scenario from $4.5\,\sigma$ to $4\,\sigma$ for the analysis based on the 3 large-recoil bins of $B\to K^*\mu^+\mu^-$ (multiplying the same uncertainties by 6 decreases the significance down to $3\,\sigma$).

If our uncertainties stemming from
Refs.~\cite{0106067,1202.4266,1303.5794} are fairly standard, one can find alternative estimates for instance in Ref.~\cite{camalich}, where larger uncertainties for $P_i'$ at large recoil are obtained due to significantly larger power-suppressed corrections. The size of these power corrections and their $q^2$-dependence was obtained from the spread between different estimates of the form factors (several sum-rule computations in different settings and Dyson-Schwinger equations). If this approach is certainly conservative, it mixes frameworks with very different levels of accuracy, $q^2$ ranges of validity, and correlations between the various form factors. Indeed, as noticed in Ref.~\cite{camalich}, the form factors are estimated in a conventional basis which is different from  the helicity basis needed for the computation of the transversity amplitudes. 
Our lack of knowledge concerning the correlations among these form factors increases the uncertainties on the helicity form factors very significantly. In our mind, averaging different parametrisations (some of them not being updated to their latest values) given in this conventional basis tends to overestimate the power-suppressed corrections in the helicity form factors. However, we agree with Ref.~\cite{camalich} that a first-principle computation of the helicity form factors would be the best way to improve the accuracy on this type of systematics.

One could also consider the alternative approach of naive factorisation following, e.g., Ref.~\cite{buras}: the \mbox{$B\to K^*\mu^+\mu^-$} form factors are then treated as independent and not reduced to two form factors $\xi_{\perp,||}$, and contributions to the transversity amplitudes from hard-gluon interactions (in particular, spectator interaction) are neglected. As shown in Ref.~\cite{buras}, these two approaches yield slightly shifted results for the $q^2$ dependence of some observables, but they agree on their qualitative variations. 
We have performed fits identical to the ones
presented in Sec.~\ref{sec:C7C9}, with very similar results [preference for $\dC9<0$, SM hypothesis with a large pull but less significance for the large-recoil data], indicating that our results are robust with respect to the use of naive or NLO QCD factorisation.

\section*{CONCLUSIONS AND PERSPECTIVES}

We have combined the recent LHCb measurements of $B\to K^*\mu^+\mu^-$ observables~\cite{Aaij:2013iag,LHCbPprime} with other radiative modes in a fit to Wilson coefficients, using the framework of our previous works~\cite{1207.2753,1303.5794}. We have found a strong indication for a negative NP contribution to the coefficient  $\C{9}$,
at $4.5 \,\sigma$ using large-recoil data ($3.9 \,\sigma$ using both large- and low-recoil data). Our results 
 correspond to $\C{9}$ inside a 68 \% C.L. range $2.2\leq \C{9} \leq 2.8$ to be compared with $\C{9}^{\rm SM}=4.07$ at the scale $\mu_b=4.8$ GeV. This is the main conclusion of our analysis of LHCb $B\to K^*\mu^+\mu^-$ measurements.

\begin{figure}
\includegraphics[width=8.3cm]{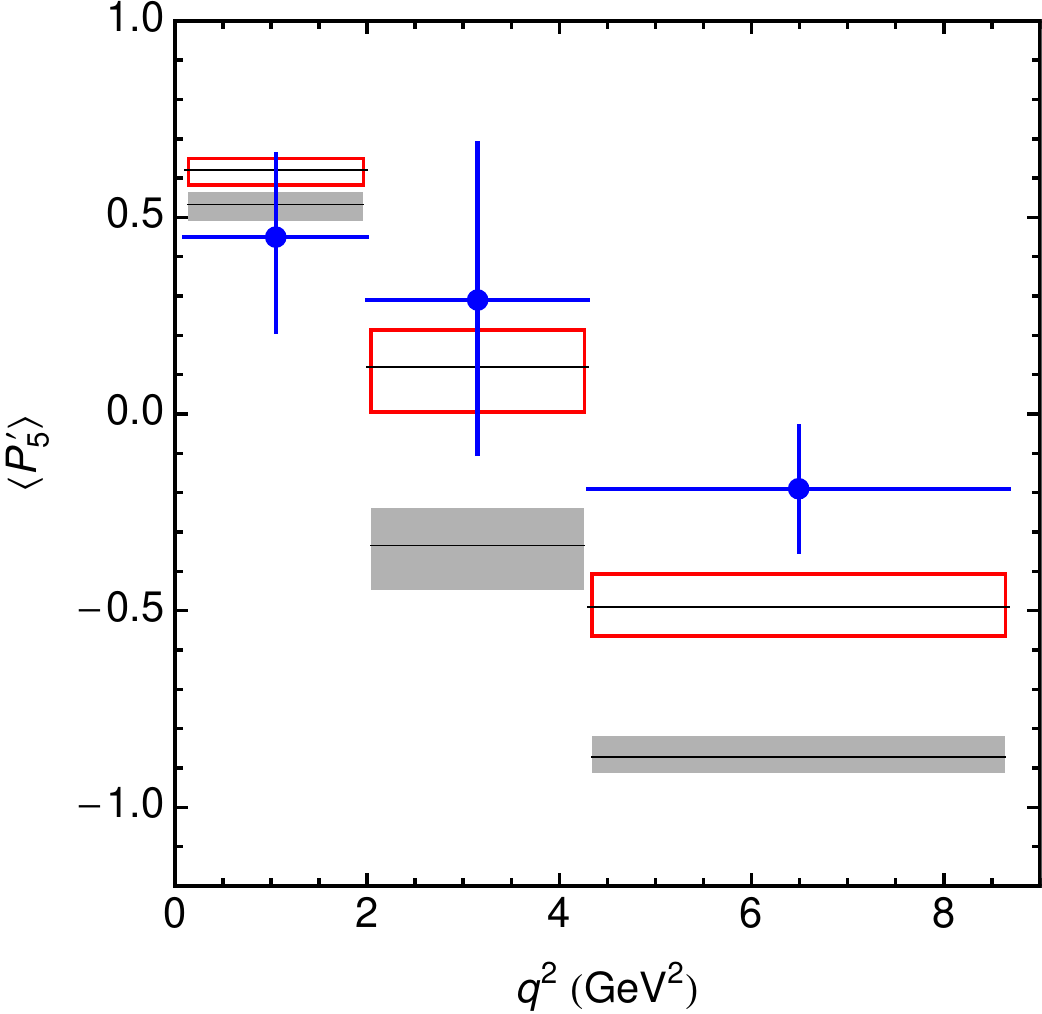}
\caption{Improvement in the $q^2$-dependence of $P_5^\prime$ in the illustrative case $\dC9=\dC{9'}=-1.5$ (and NP contributions to the other Wilson coefficients set to zero).}
\label{fig:P2C9C9prime}
\end{figure}

\begin{figure*}
\includegraphics[width=8.2cm,height=5cm]{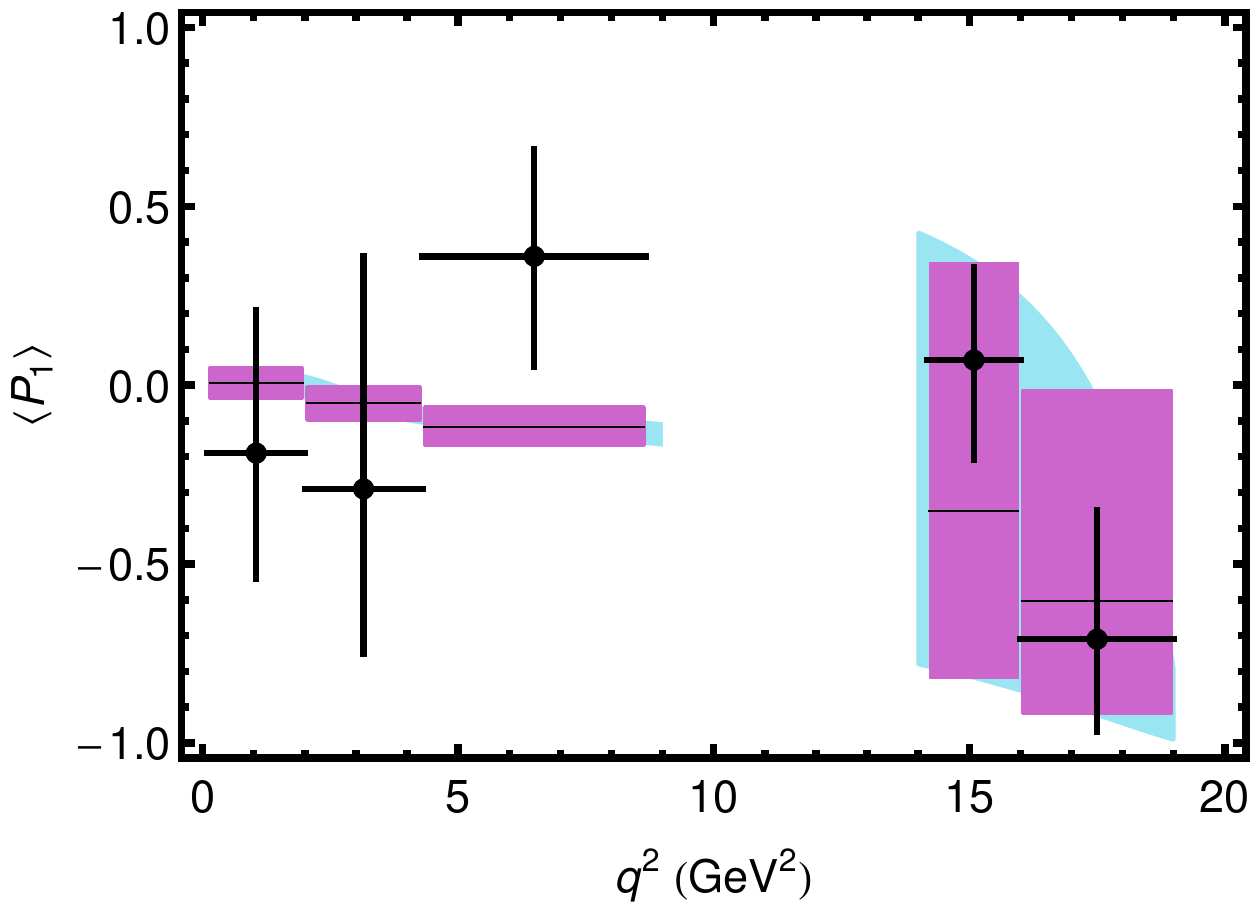}\hspace{0.6cm}
\includegraphics[width=8.2cm,height=5cm]{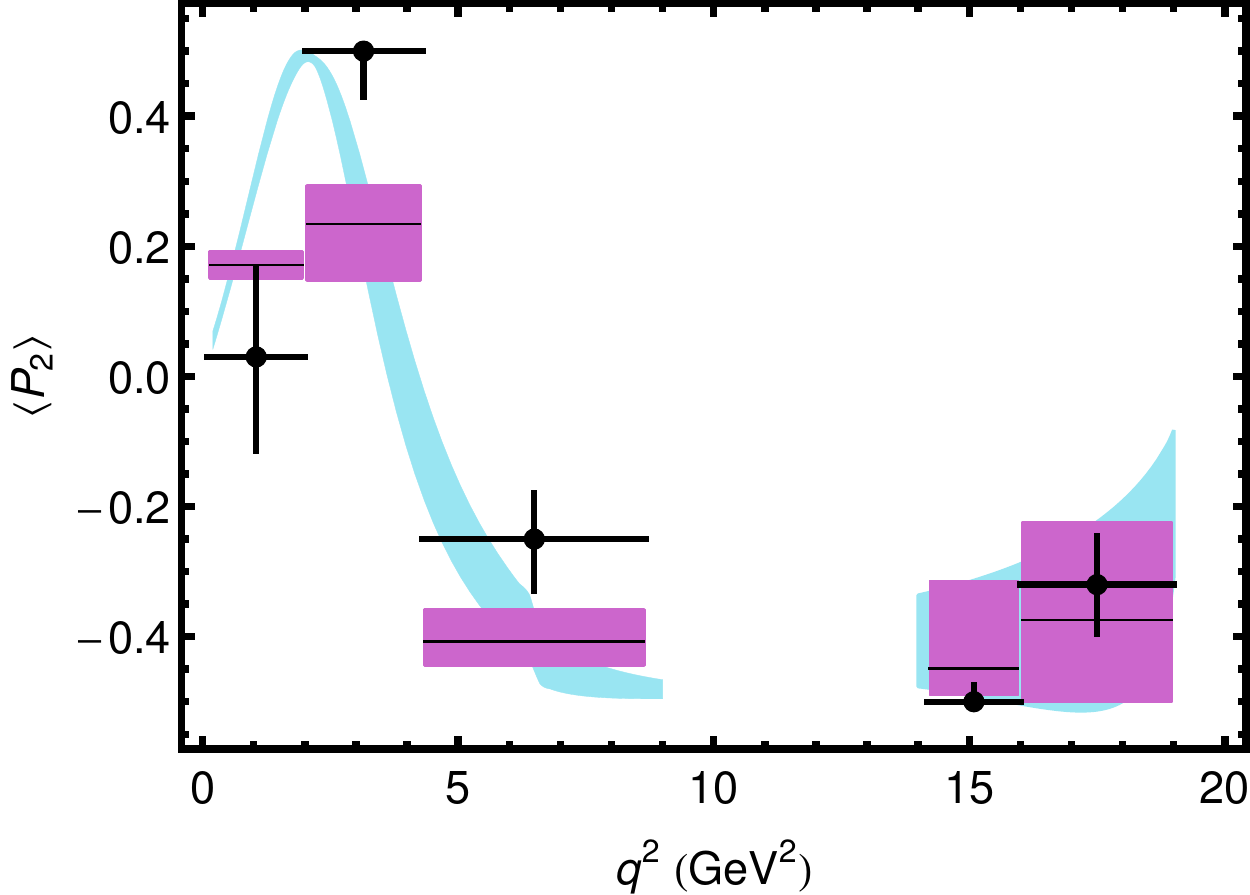}\\[2mm]
\includegraphics[width=8.2cm,height=5cm]{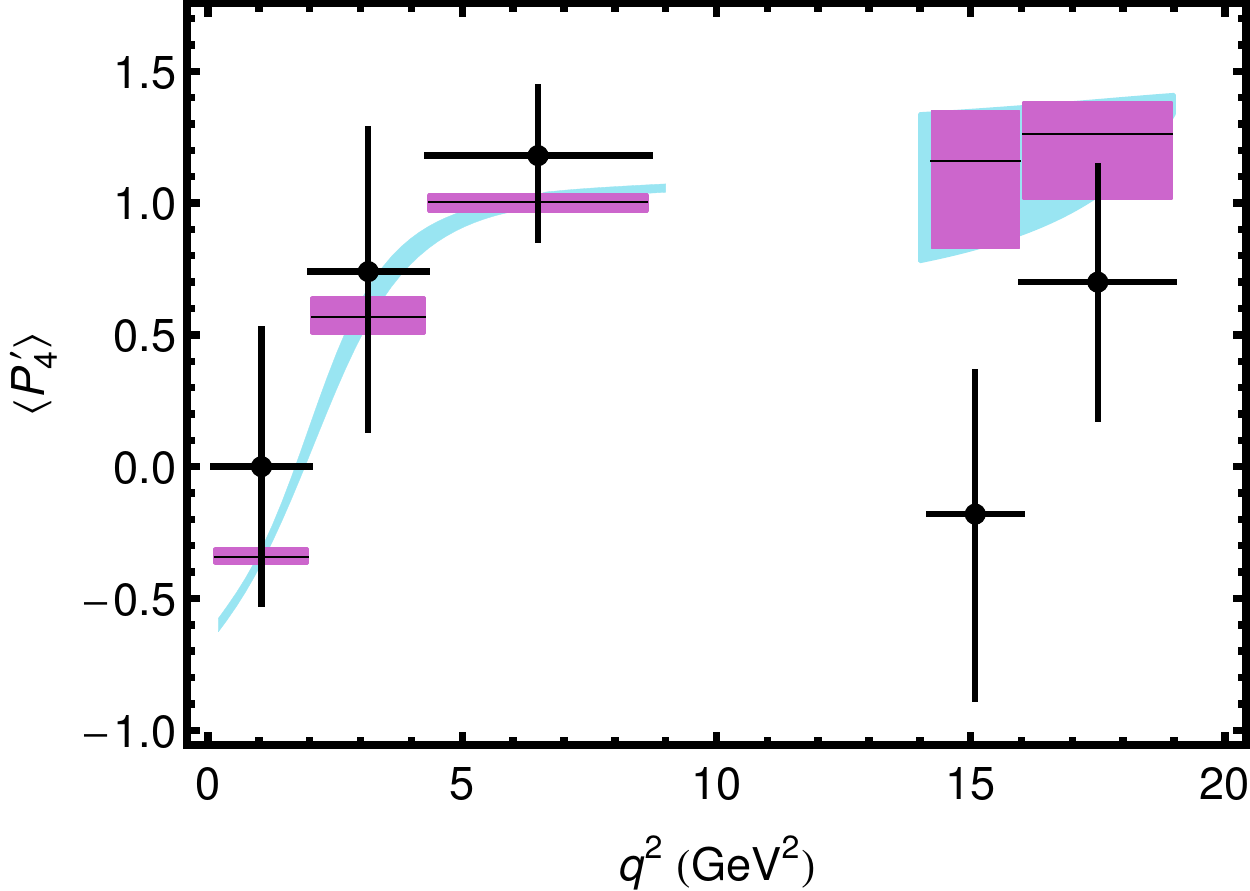}\hspace{0.6cm}
\includegraphics[width=8.2cm,height=5cm]{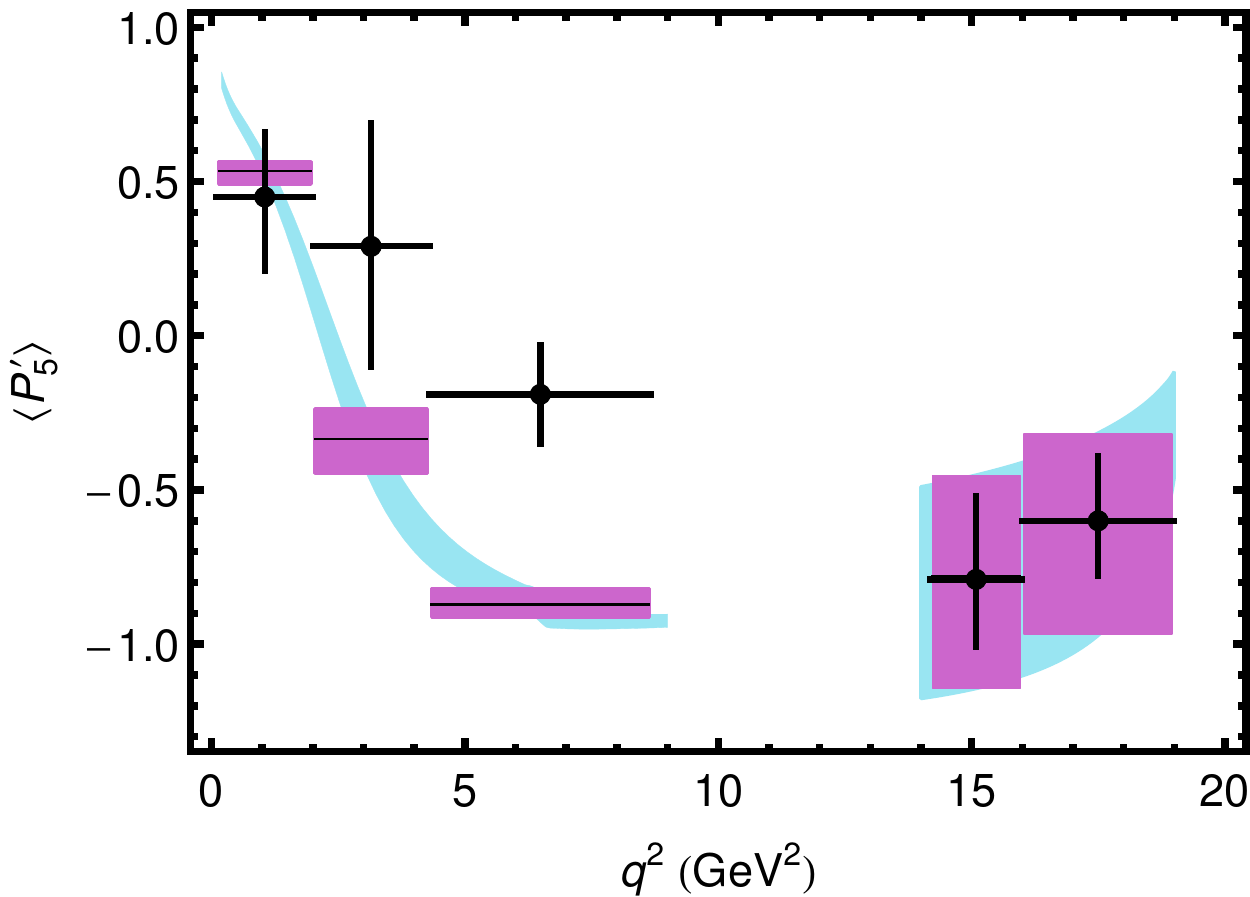}\\[2mm]
\includegraphics[width=8.2cm,height=5cm]{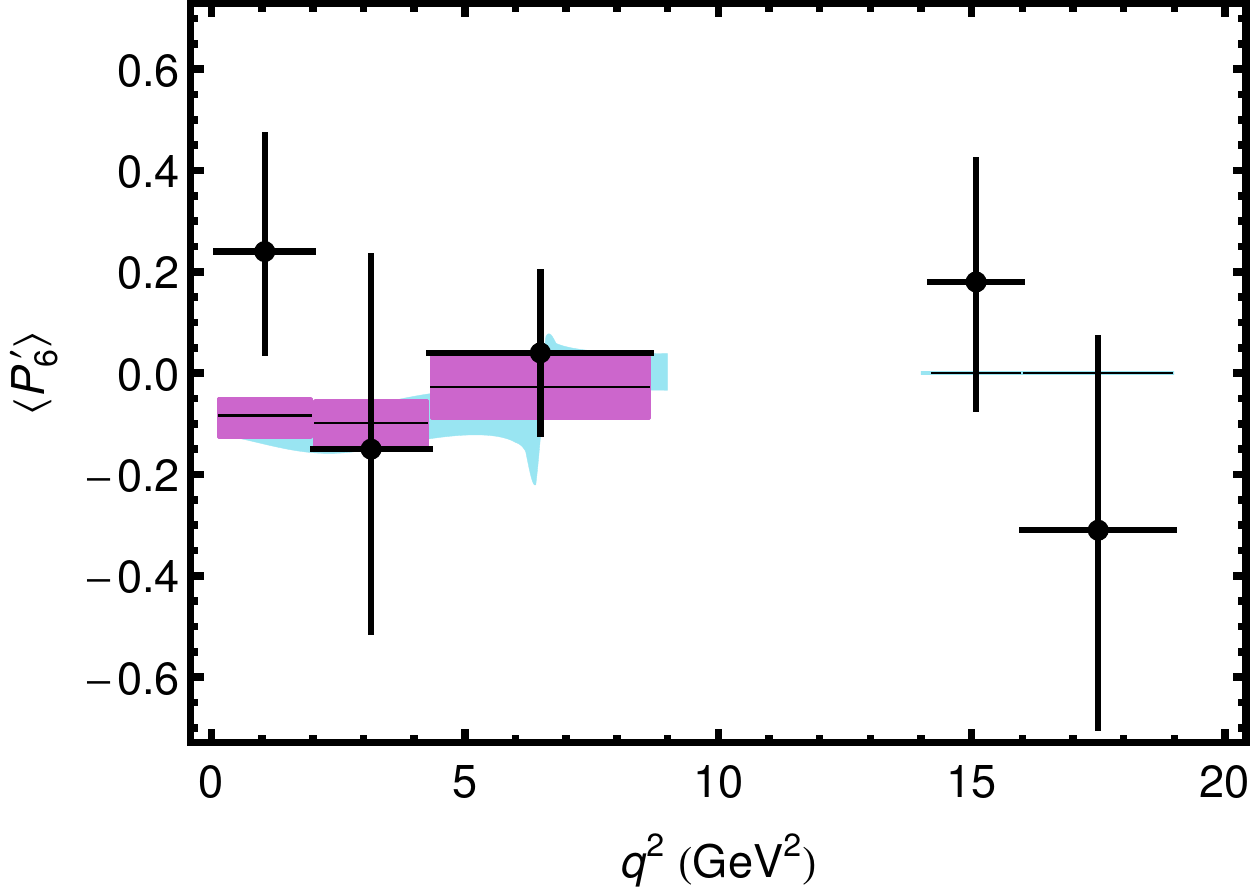}\hspace{0.6cm}
\includegraphics[width=8.2cm,height=5cm]{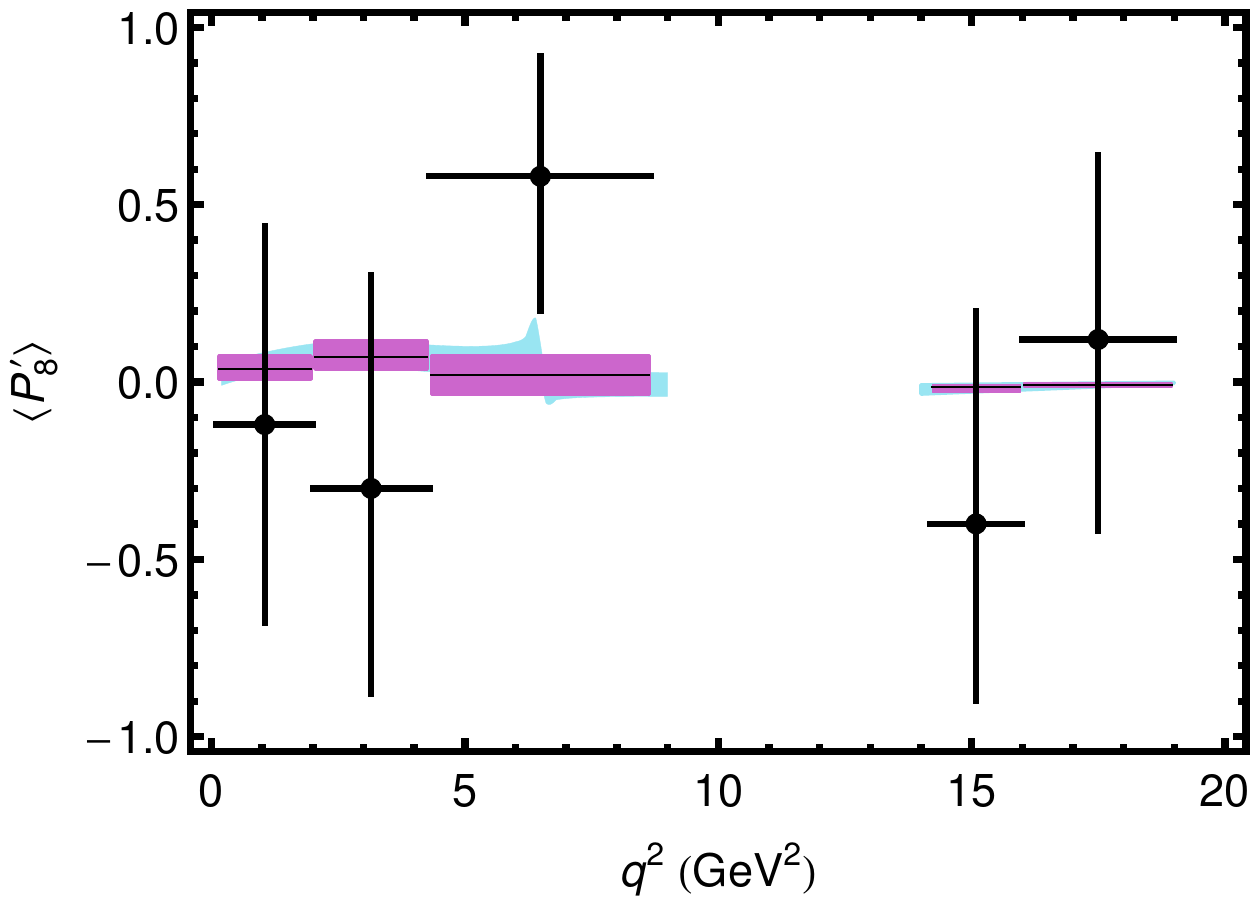}\\[2mm]
\includegraphics[width=8.2cm,height=5cm]{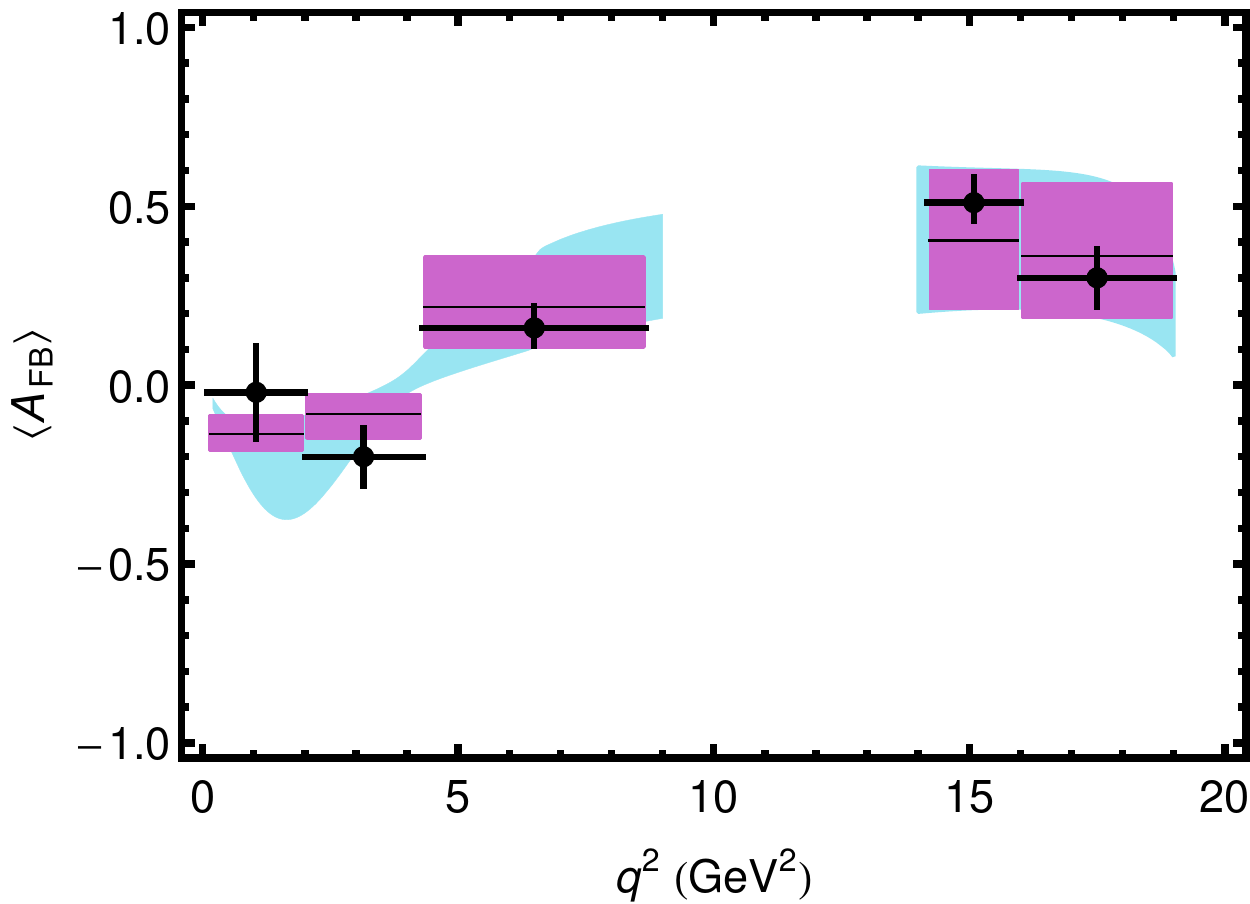}\hspace{1cm}
\includegraphics[width=7.8cm,height=5cm]{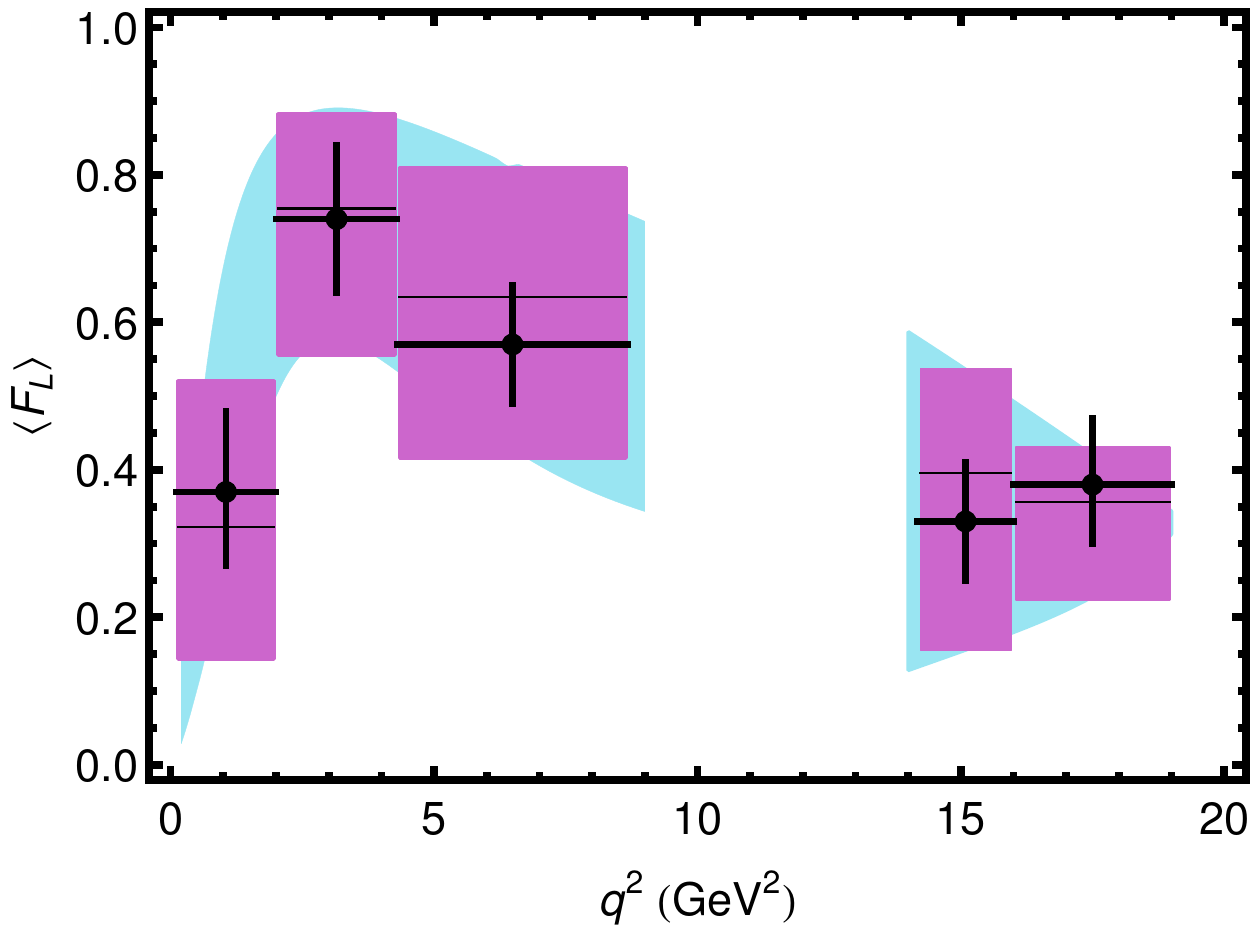}
\caption{Experimental measurements and SM predictions for some $B\to K^*\mu^+\mu^-$ observables. The black crosses are the experimental LHCb data. The blue band corresponds to the SM predictions for the differential quantities, whereas the purple boxes indicate the corresponding binned observables.}
\label{figsm}
\end{figure*}

We also observe a slight preference for negative values in $\dC{7}$ (mainly driven by radiative constraints), but no clear-cut evidence for $\dC{7',9',10'} \neq 0$.
The situation for $\C{7^\prime}$ could be clarified with a substantial reduction of error bars in $P_1$, whereas
the case for $\dC{9'} <0 $ is supported by the data at large recoil, but not at low recoil (an illustration of the improvement brought by $\dC{9'} <0 $ at large recoil is shown in Fig.~\ref{fig:P2C9C9prime}).

We emphasise that the same mechanism $\dC9<0$ ($\dC7 <0$) is remarkably efficient and economical in explaining
the whole pattern of deviations indicated by the recent LHCb results~\cite{Aaij:2013iag,LHCbPprime}: small discrepancies in $\afb$, tensions in $P_2$ and $P_5'$, preference for a higher position of the zero of $\afb$ and $P_2$.

If this pattern is confirmed, it remains to determine what kind of NP could accommodate a situation where $\dC{9}<0$ is large, with the possibility of additional contributions to $\dC{7}$ and $\dC{9'}$. 
A natural possibility would consist in a $Z'$ gauge boson~\cite{Langacker:2008yv}, coupling to left-handed quarks only but equally to left- and right-handed charged leptons, and with flavour-changing couplings to down-type quarks. This category of models and the flavour constraints set on them were discussed in Ref.~\cite{Buras:2012jb}, even though the case discussed here (i.e., in their notation, $\Delta_L^{\mu\mu}=\Delta_R^{\mu\mu}$, $\Delta_R^{sb}\simeq 0$ and $\Delta_L^{sb}$ with the same phase as $V_{tb}V_{ts}^*$) was not considered specifically in this reference. In the minimal scenario,
the $Z'$ gauge boson would contribute to $\Delta m_s$, but not to the $B_s-\bar{B}_s$ mixing phase. As an illustration, such a boson with $M_{Z'}=1$ TeV could yield $\dC9 \sim -1.5$ and be compatible with $\Delta m_s$ (using
the values in Ref.~\cite{Buras:2012jb}), provided that its coupling to muons $\Delta_L^{\mu\mu}=\Delta_R^{\mu\mu}$ is of order 0.1.
It would be worth checking if such a scenario could be framed in a specific model showing a good agreement with all the available $\Delta F=2$ measurements.

Obviously, this analysis is only a starting point, illustrating the potential of $B\to K^*\mu^+\mu^-$ observables to unveil interesting patterns of NP through rare radiative decays, and many improvements and confirmations should be gathered before reaching a definitive conclusion on the situation discussed here.

The main improvement for the analysis concerns the reduction of uncertainties. On the experimental side, the LHCb results \cite{Aaij:2013iag,LHCbPprime} are based on 1/fb of data collected in 2011.  Adding the 2012 data (another 2/fb already on disk) will constitute a big improvement concerning statistical uncertainties. In addition, one should be able to improve on some of the primary observables $P_i$, as well as on the determination of the S-wave pollution, by using new foldings (see App.~\ref{app:swave}). In order to avoid the potential pollution of $\C9$ from charm-loop effects, it is essential that the LHCb experiment provides future results for $B\to K^*\mu^+\mu^-$ with a finer $q^2$ binning, with several narrow bins between 1 and 6 GeV$^2$, and a summary of the correlations among the various measurements. 
On the theoretical side, since $\C{9}$ seems to be the main Wilson coefficient affected by NP, charm-loop effects become a very important issue, with several questions left open. It would be very useful for the theorists to converge on the scheme and the scale of the perturbative charm-quark contribution, as well as to provide alternative and/or improved estimates of the long-distance contribution obtained in Ref.~\cite{1006.4945}, in particular above the charm threshold. Another significant source of uncertainties comes from the form factors, for which new lattice results should bring more control on the low recoil region~\cite{mescia,Liu:2011raa}. In order to decrease the uncertainty attached to the form factors, it will also become essential that their estimates are provided including correlations, or in the basis of helicity form factors discussed in Refs.~\cite{Bharucha:2010im,camalich}.

An essential aspect consists in cross-checking and confirming the results from $B\to K^*\mu^+\mu^-$ on $\dC9$ through other channels accessible to LHCb and with good prospects of improving on our knowledge of the form factors. The $B\to K\mu^+\mu^-$ decay~\cite{Aaij:2012vr} gives a linear constraint between $\C{7}$ and $\C{9}$ involving pseudoscalar-to-pseudoscalar form factors well suited for lattice simulations~\cite{Becirevic:2012fy,mescia,Zhou:2011be}. The $B_s\to\phi \mu^+\mu^-$~\cite{Aaij:2013aln} has the same potential as $B\to K^*\mu^+\mu^-$ in terms of angular observables, with the added interest of a very narrow $\phi$ final state, avoiding the difficult simulation of wide resonances on the lattice. Finally, the $\Lambda_b\to\Lambda \mu^+\mu^-$ decay~\cite{Aaltonen:2011qs, Aaij:2013hna} is also a useful cross-check, with a different angular structure~\cite{Gutsche:2013pp} and very recent estimates for the relevant form factors~\cite{Feldmann:2011xf, Wang:2008sm,Wang:2009hra,Detmold:2012vy}. 

In all these aspects, a deep interplay between experimental and theoretical analyses will prove essential to confirm the pattern of NP suggested by the \mbox{$B\to K^*\mu^+\mu^-$} anomaly.

\bigskip

\begin{acknowledgments}
We would like to thank Nicola Serra, Tim Gershon and Ulrik Egede for valuable comments on the first version of the paper.
J.M. acknowledges financial support from FPA2011-25948, SGR2009-00894.
S.D.-G. thanks UAB for its kind hospitality, where part of this project has been completed.
J.V. is supported in part by ICREA-Academia funds and FPA2011-25948.
\end{acknowledgments}

\appendix

\section{Taming S-wave pollution}\label{app:swave}

In this section we focus on the S-wave pollution coming from the companion decay $B \to K_0^* \mu^+\mu^-$ ~\cite{1111.1513, 1207.4004,1209.1525} with a
 threefold aim. First, we provide the explicit form of the S-wave polluting terms entering the  different foldings allowing one to extract the primary observables. This might help in estimating the systematics if the bounds found in Ref.~\cite{1303.5794} are used. Second, we present three new foldings to disentangle the contributions from $P_{1}$ alone, $P_{1,2}$ and $P_{1,3}$. Separating $P_1$ from the other observables can be useful  to  sharpen the size of its error bars. 
 Third, we show that certain combinations of folded distributions can be sensitive to the interesting observables free from S-wave pollution. This could be combined with other approaches to the S-wave contribution\footnote{See Ref.~\cite{Doring:2013wka} for an approach to the S-wave problem using chiral perturbation theory and dispersion relations and Ref.~\cite{1210.5279} for an experimental analysis of the S-wave impact on $B\to K^*\mu^+\mu^-$.} to reduce systematic uncertainties of the experimental analysis.

 \begin{table*}[!ht] 
\ra{1.15}
\rb{2.5mm}
\begin{tabular}{@{}cccc@{}}
\toprule[1.2pt]
Obs. & S-wave & Folding & $\hat\phi$ range\\
\midrule[1.1pt]\\[-3.5mm]
$P_{1,2,3}$ & $A_s$ & 
$d\Gamma(\hat\phi, \hat\theta_l, \hat\theta_K) + d\Gamma(\hat\phi-\pi, \hat\theta_l, \hat\theta_K)$ & $[0,\pi]$\\

$P_1$ & $A_{s5},A_{s8}$ & $d\Gamma(\hat\phi, \hat\theta_l, \hat\theta_K) + d\Gamma(\hat\phi, \hat\theta_l, \pi - \hat\theta_K) + d\Gamma(-\hat\phi, \pi - \hat\theta_l, \hat\theta_K) + d\Gamma(-\hat\phi, \pi - \hat\theta_l, \pi - \hat\theta_K)$ & $[0,\pi]$\\
$P_1$ and $P_2$ & $A_{s4},A_{s5}$ & $d\Gamma(\hat\phi, \hat\theta_l, \hat\theta_K) + d\Gamma(\hat\phi, \hat\theta_l, \pi - \hat\theta_K) + d\Gamma(-\hat\phi, \hat\theta_l, \hat\theta_K) +  d\Gamma(-\hat\phi, \hat\theta_l, \pi - \hat\theta_K)$ & $[0,\pi]$\\
$P_1$ and $P_3$ & $A_{s5},A_{s7}$ & $d\Gamma(\hat\phi, \hat\theta_l, \hat\theta_K) + d\Gamma(\hat\phi, \hat\theta_l, \pi - \hat\theta_K) + d\Gamma(\hat\phi, \pi - \hat\theta_l, \hat\theta_K) + d\Gamma(\hat\phi, \pi - \hat\theta_l, \pi - \hat\theta_K)$ & $[0,\pi]$\\
$P_1$ and $P_4'$ & $A_{s5}$ & $d\Gamma(\hat\phi, \hat\theta_l, \hat\theta_K) + d\Gamma(-\hat\phi, \hat\theta_l, \hat\theta_K) + d\Gamma(\hat\phi, \pi - \hat\theta_l, \pi - \hat\theta_K) + d\Gamma(-\hat\phi, \pi - \hat\theta_l, \pi - \hat\theta_K)$ & $[0,\pi]$\\
$P_1$ and $P_5'$ & $A_{s},A_{s5}$ & $d\Gamma(\hat\phi, \hat\theta_l, \hat\theta_K) + d\Gamma(-\hat\phi, \hat\theta_l, \hat\theta_K) + d\Gamma(\hat\phi, \pi - \hat\theta_l, \hat\theta_K) +  d\Gamma(-\hat\phi, \pi - \hat\theta_l, \hat\theta_K)$ & $[0,\pi]$\\
$P_1$ and $P_6'$ & $A_{s},A_{s7}$ & $d\Gamma(\hat\phi, \hat\theta_l, \hat\theta_K) + d\Gamma(\pi - \hat\phi, \hat\theta_l, \hat\theta_K) + d\Gamma(\hat\phi, \pi - \hat\theta_l, \hat\theta_K) +  d\Gamma(\pi - \hat\phi, \pi - \hat\theta_l, \hat\theta_K)$& $[-\pi/2,\pi/2]$\\
$P_1$ and $P_8'$ & $A_{s7}$ & $d\Gamma(\hat\phi, \hat\theta_l, \hat\theta_K) + d\Gamma(\pi - \hat\phi, \hat\theta_l, \hat\theta_K) + d\Gamma(\hat\phi, \pi - \hat\theta_l, \pi - \hat\theta_K) + d\Gamma(\pi - \hat\phi, \pi - \hat\theta_l, \pi - \hat\theta_K)$ & $[-\pi/2,\pi/2]$\\
\bottomrule[1.2pt]
\end{tabular}
\caption{Foldings needed to single out the interesting observables, with the corresponding remaining S-wave pollution. For all foldings, $\hat\theta_\ell$ and $\hat\theta_K$ lie within $[0,\pi/2]$, whereas $\hat\phi$ has different ranges depending on the observables considered.}\label{foldings}
\end{table*}

Our starting point is the distribution given in Eq.~(43) of Ref.~\cite{1303.5794} using the definitions for the angular variables defined there and working in the massless lepton limit (see Ref.~\cite{1209.1525} for the massive case).
We  determine the S-wave pollution accompanying  the relevant folded distributions that can be used to extract the primary observables:
 \begin{itemize}
 \item $P_{1,2,3}$: Identifying $\phi \leftrightarrow \phi + \pi $  (when $\phi<0$) amounts to the folded distribution  $d\hat \Gamma=d\Gamma(\hat \phi)+ d\Gamma(\hat\phi-\pi)$. Once normalised to the full distribution the result (see Sec.7 of \cite{1303.5794} for definitions) is $d\hat\Gamma/\Gamma_{\rm full}=
\frac{x}{2} + \frac{9}{16 \pi} F_T \sin^2\hat\theta_K ( 2 P_2 \cos\hat\theta_\ell - P_3 \sin 2 \hat\phi \sin^2\hat\theta_\ell)(1-F_S)+\delta^1_{sw}$
  where $x=\frac{9}{32 \pi} ( 8 F_L \cos^2 \hat\theta_K\sin^2 \hat\theta_\ell+ F_T \sin^2\hat\theta_K (3 + \cos 2 \hat\theta_\ell) + 2 F_T P_1 \sin^2\hat\theta_K \cos 2 \hat\phi \sin^2 \hat\theta_\ell) (1-F_S)$ and    $\delta^{(1)}_{sw}=
  \frac{3}{8\pi} (F_S + A_S \cos \hat\theta_K)\sin^2 \hat\theta_\ell $ stands for the S-wave contribution. The folded angle $\hat\phi$ is defined in the range $\hat\phi\in [0,\pi]$  and all other angles are inside their usual range $\hat\theta_l,\hat\theta_K\in[0,\pi]$. Other foldings providing subsets of $P_{1,2,3}$ can be found in Table~\ref{foldings}.

 \item $P_4^\prime$: A double folded distribution is necessary $d\hat\Gamma=d\Gamma(\hat \phi)+ d\Gamma(-\hat\phi)+d\Gamma(\pi -\hat\phi, \pi-\hat\theta_\ell)+d\Gamma(\hat\phi-\pi,\pi-\hat\theta_\ell)$ and once normalised $d\hat\Gamma/\Gamma_{\rm full}=x + \frac{9}{16 \pi} \sqrt{F_T F_L} P_4^\prime \cos \hat\phi\sin 2 \hat\theta_K \sin2\hat\theta_\ell (1-F_S) + \delta^{(2)}_{sw}$ where $\delta^{(2)}_{sw}=2  \delta^{(1)}_{sw} + \frac{3}{4\pi}  A_S^4  \cos\hat\phi \sin\hat\theta_K\sin 2\hat \theta_\ell$ with  $\hat\phi\in[0,\pi]$,  $\hat\theta_\ell\in[0,\pi/2]$, $\hat\theta_K\in[0,\pi]$. Another possible folding leading to $P_4^\prime$ can be found in Table~\ref{foldings}. 
  
 \item $P_5^\prime$: A double folded distribution is also required $d\hat\Gamma=d\Gamma(\hat \phi)+ d\Gamma(-\hat\phi)+d\Gamma(\hat\phi,\pi-\hat\theta_\ell)+d\Gamma(-\hat\phi,\pi-\hat\theta_\ell)$ leading to $d\hat\Gamma/\Gamma_{\rm full}=x + \frac{9}{8 \pi} \sqrt{F_T F_L} P_5^\prime \cos\hat \phi\sin 2 \hat\theta_K \sin\hat\theta_\ell (1-F_S) + \delta^{(3)}_{sw}$ where $\delta^{(3)}_{sw}=2  \delta^{(1)}_{sw} + \frac{3}{4\pi}   A_S^5  \cos\hat\phi \sin\hat\theta_K\sin \hat \theta_\ell$. Here $\hat\phi\in[0,\pi]$,  $\hat\theta_\ell\in[0,\pi/2]$ and $\hat\theta_K\in[0,\pi]$.
 
 \end{itemize}

The explicit form of the folding  in terms of distributions for the last two primary observables $P_{6,8}^\prime$ depends on the region in parameter space ($\phi,\theta_\ell,\theta_K$) chosen.
We provide here the folded distributions for $P_6^\prime$ and $P_8^\prime$ in the region $\hat\phi\in[0,\pi/2]$ (a similar folding can be obtained for $\hat\phi\in[-\pi/2,0]$). 

\begin{itemize}

 \item $P_6^\prime$: The associated folded distribution is $d\hat\Gamma=d\Gamma(\hat \phi)+ d\Gamma(\pi-\hat\phi)+d\Gamma(\hat\phi,\pi-\hat\theta_\ell)+d\Gamma( \pi-\hat\phi  ,\pi-\hat\theta_\ell)$ corresponding to $d\hat\Gamma/\Gamma_{\rm full}=x - \frac{9}{8 \pi} \sqrt{F_T F_L} P_6^\prime \sin\hat \phi\sin 2 \hat\theta_K \sin\hat\theta_\ell (1-F_S) + \delta^{(4)}_{sw}$ where $\delta^{(4)}_{sw}=2  \delta^{(1)}_{sw} + \frac{3}{4\pi}  A_S^7  \sin\hat\phi \sin\hat\theta_K\sin \hat \theta_\ell$ and $\hat\theta_\ell\in[0,\pi/2]$, $\hat\theta_K\in[0,\pi]$.
 
 \item $P_8^\prime$: The folded distribution is in this case $d\hat\Gamma = d\Gamma(\hat \phi) + d\Gamma(\pi-\hat\phi) + d\Gamma(\pi-\hat\theta_\ell,\pi-\hat\theta_K) + d\Gamma(\pi-\hat\phi, \pi-\hat\theta_\ell,\pi-\hat\theta_K) = x + \frac{9}{16 \pi} \sqrt{F_T F_L} P_8^\prime \sin \hat\phi\sin 2 \hat\theta_K \sin2\hat\theta_\ell (1-F_S) + \delta^5_{sw}$ where $\delta^5_{sw}=\frac{3}{4\pi} F_S \sin^2\hat\theta_\ell + \frac{3}{4\pi}  A_S^7  \sin\hat\phi \sin\hat\theta_K\sin  \hat\theta_\ell$ with $\hat\theta_\ell\in[0,\pi/2]$, $\hat\theta_K\in[0,\pi]$.
 
 \end{itemize}

We summarise  this discussion in Table~\ref{foldings} presenting some of the foldings already discussed, and other possibilities, with their sensitivity to primary observables and S-wave polluting terms (besides $F_S$). 

Given that the present statistics does not allow one to fit all S-wave coefficients, we suggest combinations of folded distributions sensitive to the interesting observables $P_{2}$ and $P_{4,5}^\prime$ free from S-wave pollution (besides a global $1-F_S$ factor), but at the price of reducing the experimental sensitivity.
We provide in the following two examples of this approach, one for $P_2$ and one for $P_{4,5}^\prime$.

 First, it was found in Ref.~\cite{1209.1525} that the identification $\phi \leftrightarrow \phi+\pi$ and $\theta_\ell \leftrightarrow \pi-\theta_\ell$ leads to a folding similar to the first folding in Table~\ref{foldings} but with twice the S-wave term. In this way the combination of distributions $d\hat\Gamma = d\Gamma(\hat \phi,\pi-\hat\theta_\ell) + d\Gamma(\hat\phi-\pi,\pi-\hat\theta_\ell) - d\Gamma(\hat\phi) - d\Gamma(\hat\phi-\pi)$ once normalised allows  a direct measurement of $P_2$, i.e., $d\hat\Gamma/\Gamma_{\rm full}=-\frac{9}{4\pi} F_T P_2 \cos\hat\theta_\ell \sin^2\hat\theta_K (1-F_S)$.
 
 The second example combines the  two distributions for $P_{4,5}^\prime$  described at the beginning of this Section
 with three distributions given by Eqs.~(28), (41) and (42)  in Ref.\cite{1209.1525}. 
 In this case the resulting combination is $d\hat\Gamma =
 d\Gamma(-\hat\phi)+d\Gamma(-\hat\phi+\pi,\pi-\hat\theta_\ell)+
 d\Gamma(\hat\phi-\pi,\pi-\hat\theta_K)+d\Gamma(-\hat\phi,\pi-\hat\theta_\ell)-d\Gamma(\hat\phi-\pi)-d\Gamma(-\hat\phi,\pi-\hat\theta_K)-d\Gamma(\pi-\hat\theta_\ell)-d\Gamma(\hat\phi-\pi,\pi-\hat\theta_\ell)$
  leading to $d\hat\Gamma/\Gamma_{\rm full}=\frac{9}{4\pi}\sin\hat\theta_K \cos\hat\phi\sin\hat\theta_\ell (
 \sqrt{F_L F_T} (P_5^\prime +P_4^\prime \cos\hat\theta_\ell)\cos\hat\theta_K + F_T P_3 \sin\hat\phi\sin\hat\theta_K\sin\hat\theta_\ell) (1-F_S)$.

Finally, we would like to point out that one can also tame the S-wave contribution using the bounds on the $A_S^i$ coefficients presented in Ref.~\cite{1303.5794}, which however require
a measurement of $F_S$. This can be achieved through the folded distribution $d\hat\Gamma =
 d\Gamma(\hat\phi)+d\Gamma(\pi-\hat\theta_K)+d\Gamma(\hat\phi-\pi)+d\Gamma(\hat\phi-\pi,\pi-\hat\theta_K)$
where only $F_S$ enters as an S-wave pollution
   $d\hat\Gamma/\Gamma_{\rm full}=x+\frac{1}{4\pi} [  9 F_T \sin^2\hat\theta_K (P_2 \cos\hat\theta_\ell - P_3 \cos\hat\phi\sin\hat\phi\sin^2\hat\theta_\ell)](1-F_S)+\frac{3}{4\pi} F_S \sin^2\hat\theta_\ell  $.


\begin{thebibliography}{99}


 \bibitem{kruger} 
  F.~Kruger and J.~Matias,
  Phys.\ Rev.\ D\ {\bf 71}, 094009  (2005)
  [hep-ph/0502060].
 
 
 \bibitem{matias1} 
  U.~Egede, T.~Hurth, J.~Matias, M.~Ramon and W.~Reece,
  JHEP\ {\bf 0811}, 032  (2008)
  [arXiv:0807.2589 [hep-ph]].

 \bibitem{Lunghi:2006hc}
  E.~Lunghi and J.~Matias,
  JHEP {\bf 0704} (2007) 058
  [hep-ph/0612166].

 \bibitem{buras} 
  W.~Altmannshofer, P.~Ball, A.~Bharucha, A.~J.~Buras, D.~M.~Straub and M.~Wick,
  JHEP\ {\bf 0901}, 019  (2009)
  [arXiv:0811.1214 [hep-ph]].

\bibitem{matias2} 
  U.~Egede, T.~Hurth, J.~Matias, M.~Ramon and W.~Reece,
  JHEP\ {\bf 1010}, 056  (2010)
  [arXiv:1005.0571 [hep-ph]].

\bibitem{becirevic} 
  D.~Becirevic, E.~Schneider,
  Nucl.\ Phys.\ B\ {\bf 854}, 321  (2012)
  [arXiv:1106.3283 [hep-ph]].

\bibitem{1202.4266} 
  J.~Matias, F.~Mescia, M.~Ramon and J.~Virto,
  JHEP {\bf 1204}, 104 (2012)
  [arXiv:1202.4266 [hep-ph]].


 \bibitem{bobeth} 
  C.~Bobeth, G.~Hiller and D.~van Dyk,
  JHEP\ {\bf 1007}, 098  (2010)
  [arXiv:1006.5013 [hep-ph]].
 
 \bibitem{tensors} 
  C.~Bobeth, G.~Hiller and D.~van Dyk,
  arXiv:1212.2321 [hep-ph].
  
\bibitem{Alok:2010zd} 
  A.~K.~Alok, A.~Datta, A.~Dighe, M.~Duraisamy, D.~Ghosh and D.~London,
  JHEP {\bf 1111}, 121 (2011)
  [arXiv:1008.2367 [hep-ph]].
  
\bibitem{Alok:2011gv} 
  A.~K.~Alok, A.~Datta, A.~Dighe, M.~Duraisamy, D.~Ghosh and D.~London,
  JHEP {\bf 1111}, 122 (2011)
  [arXiv:1103.5344 [hep-ph]].
  
  \bibitem{Altmannshofer:2011gn} 
  W.~Altmannshofer, P.~Paradisi and D.~M.~Straub,
  JHEP {\bf 1204}, 008 (2012)
  [arXiv:1111.1257 [hep-ph]].
  
  \bibitem{Altmannshofer:2012az} 
  W.~Altmannshofer and D.~M.~Straub,
  JHEP {\bf 1208}, 121 (2012)
  [arXiv:1206.0273 [hep-ph]].
  
  \bibitem{Beaujean:2012uj} 
  F.~Beaujean, C.~Bobeth, D.~van Dyk and C.~Wacker,
  JHEP {\bf 1208}, 030 (2012)
  [arXiv:1205.1838 [hep-ph]].
  
\bibitem{1207.2753} 
  S.~Descotes-Genon, J.~Matias, M.~Ramon and J.~Virto,
  JHEP {\bf 1301}, 048 (2013)
  [arXiv:1207.2753 [hep-ph]].

  
\bibitem{1108.0695}  
 {\bf CDF} Collaboration,
  Phys.\ Rev.\ Lett.\  {\bf 108}, 081807 (2012)
  [arXiv:1108.0695 [hep-ex]].

\bibitem{Wei:2009zv}
  {\bf Belle} Collaboration,
  Phys.\ Rev.\ Lett.\  {\bf 103} (2009) 171801
  [arXiv:0904.0770 [hep-ex]].

\bibitem{Lees:2012tva}
  {\bf BaBar} Collaboration,
  Phys.\ Rev.\ D {\bf 86} (2012) 032012
  [arXiv:1204.3933 [hep-ex]].

\bibitem{Aaij:2013iag}
  {\bf LHCb} Collaboration,
  arXiv:1304.6325 [hep-ex].

\bibitem{LHCbPprime}
LHCb-PAPER-2013-037, in preparation.\\
See also: N.~Serra, \emph{Studies of electroweak penguin transitions of $b\to s\mu\mu$}, talk at the EPS-HEP Conference, Stockholm, July 2013.


\bibitem{1303.5794} 
  S.~Descotes-Genon, T.~Hurth, J.~Matias and J.~Virto,
  JHEP {\bf 1305}, 137 (2013)
  [arXiv:1303.5794 [hep-ph]].


 \bibitem{0106067} 
  M.~Beneke, T.~Feldmann and D.~Seidel,
  Nucl.\ Phys.\ B {\bf 612}, 25 (2001)
  [hep-ph/0106067].

 \bibitem{0412400} 
  M.~Beneke, T.~Feldmann and D.~Seidel,
  Eur.\ Phys.\ J.\ C {\bf 41}, 173 (2005)
  [hep-ph/0412400].

\bibitem{grinstein+pirjol} 
  B.~Grinstein and D.~Pirjol,
  Phys.\ Rev.\ D {\bf 70}, 114005 (2004)
  [hep-ph/0404250].

\bibitem{matias-Eps}
J. Matias, \emph{Optimizing the basis of $B\to K^*\ell^+\ell^-$ observables and understanding its tensions}, talk at the EPS-HEP Conference, Stockholm, July 2013.

\bibitem{1104.3342}
  S.~Descotes-Genon, D.~Ghosh, J.~Matias and M.~Ramon,
  JHEP {\bf 1106}, 099 (2011)
  [arXiv:1104.3342 [hep-ph]]. 
  
\bibitem{hfag} 
  Y.~Amhis {\it et al.}  [Heavy Flavor Averaging Group Collaboration],
  arXiv:1207.1158 [hep-ex].
  
  \bibitem{0712.3009} 
  T.~Huber, T.~Hurth, E.~Lunghi,
  Nucl.\ Phys.\ B {\bf 802}, 40 (2008)
  [arXiv:0712.3009 [hep-ph]].
  
  \bibitem{1211.2674} 
 {\bf LHCb} Collaboration,
  Phys.\ Rev.\ Lett.\  {\bf 110}, 021801 (2013)
  [arXiv:1211.2674 [Unknown]].
  
 \bibitem{LHCbBsmumu-Eps}
{\bf LHCb} Collaboration,
  arXiv:1307.5024 [hep-ex].
 
 \bibitem{CMSBsmumu-Eps} 
 {\bf CMS} Collaboration, 
  arXiv:1307.5025 [hep-ex].

 \bibitem{FlavourExp-Eps} 
S.~Hansmann-Menzemer, \emph{Experimental Results on Flavour Physics}, talk at the EPS-HEP Conference, Stockholm, July 2013.


\bibitem{1205.3422} 
  {\bf LHCb} Collaboration,
  JHEP {\bf 1207}, 133 (2012)
  [arXiv:1205.3422 [hep-ex]].
  
\bibitem{Aaij:2013pta} 
  {\bf LHCb} Collaboration,
  arXiv:1307.7595 [hep-ex].

\bibitem{Lenz:2010gu}
  A.~Lenz, U.~Nierste, J.~Charles, S.~Descotes-Genon, A.~Jantsch, C.~Kaufhold, H.~Lacker {\it et al.},
  Phys.\ Rev.\ D {\bf 83} (2011) 036004
  [arXiv:1008.1593[hep-ph]].

\bibitem{camalich} 
  S.~J\"ager and J.~M.~Camalich,
  arXiv:1212.2263 [hep-ph].

  \bibitem{1006.4945} 
  A.Khodjamirian, T.Mannel, A.A.Pivovarov, Y.M.Wang,
  JHEP {\bf 1009}, 089 (2010)
  [arXiv:1006.4945 [hep-ph]].

\bibitem{vandyk-Eps}
D. Van Dyk, \emph{Constraints on $|\Delta B|=|\Delta S|=1$ Wilson coefficients}, talk at the EPS-HEP Conference, Stockholm, July 2013.

\bibitem{0512066} 
  T.~Huber, E.~Lunghi, M.~Misiak and D.~Wyler,
  Nucl.\ Phys.\ B {\bf 740}, 105 (2006)
  [hep-ph/0512066].

\bibitem{Langacker:2008yv}
  P.~Langacker,
  Rev.\ Mod.\ Phys.\  {\bf 81} (2009) 1199
  [arXiv:0801.1345 [hep-ph]].

\bibitem{Buras:2012jb}
  A.~J.~Buras, F.~De Fazio and J.~Girrbach,
  JHEP {\bf 1302} (2013) 116
  [arXiv:1211.1896 [hep-ph]].

\bibitem{mescia} 
  D.~Becirevic, V.~Lubicz and F.~Mescia,
  Nucl.\ Phys.\ B {\bf 769}, 31 (2007)
  [hep-ph/0611295].
  
\bibitem{Liu:2011raa}
  Z.~Liu, S.~Meinel, A.~Hart, R.~R.~Horgan, E.~H.~Muller and M.~Wingate,
  arXiv:1101.2726 [hep-ph].

\bibitem{Bharucha:2010im}
  A.~Bharucha, T.~Feldmann and M.~Wick,
  JHEP {\bf 1009} (2010) 090
  [arXiv:1004.3249 [hep-ph]].
  
\bibitem{Aaij:2012vr}
  {\bf LHCb} Collaboration,
  JHEP {\bf 1302} (2013) 105
  [arXiv:1209.4284 [hep-ex]].
  
\bibitem{Becirevic:2012fy}
  D.~Becirevic, N.~Kosnik, F.~Mescia, E.~Schneider,
  Phys.\ Rev.\ D {\bf 86} (2012) 034034
  [arXiv:1205.5811 [hep-ph]].

\bibitem{Zhou:2011be}
  R.~Zhou {\it et al.}  [Fermilab Lattice \& MILC Collaborations],
  PoS LATTICE {\bf 2011} 298
  [arXiv:1111.0981 [hep-lat]].

\bibitem{Aaij:2013aln}
  {\bf LHCb} Collaboration,
  arXiv:1305.2168 [hep-ex].

\bibitem{Aaltonen:2011qs}
  {\bf CDF} Collaboration,
  Phys.\ Rev.\ Lett.\  {\bf 107} (2011) 201802
  [arXiv:1107.3753 [hep-ex]].

\bibitem{Aaij:2013hna}
  {\bf LHCb} Collaboration,
  arXiv:1306.2577 [hep-ex].

\bibitem{Gutsche:2013pp}
  T.~Gutsche, M.~A.~Ivanov, J.~G.~Korner, V.~E.~Lyubovitskij and P.~Santorelli,
  Phys.\ Rev.\ D {\bf 87} (2013) 074031
  [arXiv:1301.3737 [hep-ph]].

\bibitem{Feldmann:2011xf}
  T.~Feldmann and M.~W.~Y.~Yip,
  Phys.\ Rev.\ D {\bf 85} (2012) 014035
   [Erratum-ibid.\ D {\bf 86} (2012) 079901]
  [arXiv:1111.1844 [hep-ph]].
  
 \bibitem{Wang:2008sm} 
  Y.~-m.~Wang, Y.~Li and C.~-D.~Lu,
  Eur.\ Phys.\ J.\ C {\bf 59}, 861 (2009)
  [arXiv:0804.0648 [hep-ph]].
  
  \bibitem{Wang:2009hra} 
  Y.~-M.~Wang, Y.~-L.~Shen and C.~-D.~Lu,
  Phys.\ Rev.\ D {\bf 80}, 074012 (2009)
  [arXiv:0907.4008 [hep-ph]].
  

\bibitem{Detmold:2012vy}
  W.~Detmold, C.~-J.~D.~Lin, S.~Meinel, M.~Wingate,
  Phys.\ Rev.\ D {\bf 87} (2013) 074502
  [arXiv:1212.4827 [hep-lat]].

\bibitem{1111.1513} 
  C.~-D.~Lu and W.~Wang,
  Phys.\ Rev.\ D {\bf 85}, 034014 (2012)
  [arXiv:1111.1513 [hep-ph]].

 \bibitem{1207.4004} 
  D.~Becirevic and A.~Tayduganov,
  Nucl.\ Phys.\ B {\bf 868}, 368 (2013)
  [arXiv:1207.4004 [hep-ph]].
   
 \bibitem{1209.1525} 
  J.~Matias,
  Phys.\ Rev.\ D {\bf 86}, 094024 (2012)
  [arXiv:1209.1525 [hep-ph]].

\bibitem{Doring:2013wka}
  M.~Doring, U.~-G.~Mei{\ss}ner and W.~Wang,
  arXiv:1307.0947 [hep-ph].
  
  \bibitem{1210.5279} 
  T.~Blake, U.~Egede, A.~Shires ,
  JHEP {\bf 1303}, 027 (2013)
  [arXiv:1210.5279 [hep-ph]].

\end{thebibliography}
\end{document}